\documentclass[fleqn,usenatbib]{mnras}

\usepackage{newtxtext,newtxmath}
\usepackage[T1]{fontenc}
\usepackage[utf8]{inputenc} % Required for inputting international characters
\usepackage{subfig}
\usepackage{graphicx}

\usepackage{placeins}
\usepackage{caption}
\usepackage{subcaption}
\usepackage{xspace}

\usepackage{graphicx}	% Including figure files
\usepackage{amsmath}	% Advanced maths commands

\usepackage{mhchem}

\usepackage[autostyle=true]{csquotes} % Required to generate language-dependent quotes in the bibliography
\usepackage{soul}
\DeclareRobustCommand{\VAN}[3]{#2}
\let\VANthebibliography\thebibliography
\def\thebibliography{\DeclareRobustCommand{\VAN}[3]{##3}\VANthebibliography}

\usepackage{graphicx}	% Including figure files
\usepackage{amsmath}	% Advanced maths commands

\newcommand{\ngc}{NGC\,5643}

\newcommand{\oth}{{[\sc O \romannumeral 3]} \xspace}
\newcommand{\hii}{{H$_{2}$}\xspace}
\newcommand{\hi}{\ion{H}{i}}
\newcommand{\matHI}{{\rm H {\hskip 0.005cm \texttt{I}}}}

\newcommand{\mstar}{$M_\star$}
\newcommand{\msun}{{M$_\odot$}}

\newcommand{\kms}{$\,$km$\,$s$^{-1}$}

\title[Infalling cold gas on \ngc]{MeerKAT discovery of an infalling cold gas tail onto the nearby barred spiral galaxy, \ngc.
}

\author[K. C. Santana et al.]{
Karina C. Santana,$^{1}$
Filippo M. Maccagni,$^{2}$
Roger Deane$^{1,3}$
and Julia Healy$^{4,5}$
\\
$^{1}$Wits Centre for Astrophysics, School of Physics, University of the Witwatersrand, 1 Jan Smuts Avenue, 2000, Johannesburg, South Africa \\
$^{2}$INAF -- Osservatorio Astronomico di Cagliari, via della Scienza 5, 09047 Selargius, CA, Italy \\
$^{3}$Department of Physics, University of Pretoria, Hatfield, Pretoria, 0028, South Africa \\
$^{4}$ Jodrell Bank Centre for Astrophysics, School of Physics and Astronomy, University of Manchester, Oxford Road, Manchester M13 9PL, UK \\
$^{5}$ United Kingdom SKA Regional Centre (UKSRC), UK
}

\date{Accepted 2025 May 15. Received 2025 May 13; in original form 2024 November 22}

\pubyear{2024}

\begin{document}

\label{firstpage}
\pagerange{\pageref{firstpage}--\pageref{lastpage}}
\maketitle

\begin{abstract}
The detailed study of gas flows in local Active Galactic Nuclei (AGN) is essential for understanding the regulation of star formation and black hole growth, which are fundamental to galaxy evolution. One such AGN case study is \ngc, a nearby ($D_{L}\sim17.3$~Mpc) star-forming, late-type, Seyfert galaxy, where inflows and outflows have been observed in detail. 
\ngc\ has been studied at multiple wavelengths, however, a key missing component is sensitive, high-resolution neutral hydrogen (\hi) observations. We present 21-cm observations of \ngc\ with MeerKAT, revealing six low-\hi\ mass ($M_{\text{\hi}}\sim10^{7}$\,\msun) sources surrounding \ngc\ and \hi\ in IC\,4444, $\sim230$\,kpc north of \ngc. 
In \ngc, \hi\ extends beyond the stellar disk with several morphological and kinematical asymmetries. North of the disk is an extended 30~kpc tail with counter-rotating velocities.
This is \hi\ gas accreting onto the regularly rotating disk of \ngc\ from the environment.
Within the spiral arms of the disk, we identify extraplanar gas components, tracing galactic fountains driven by star formation regions. These fountains have a molecular gas component and show an increased \hii/\hi\ ratio.
In the circum-nuclear region, we observe spatially unresolved \hi\ absorption that is slightly blue-shifted ($\sim 72$\,\kms) with an \hi\ emission counterpart at redshifted velocities. 
These MeerKAT observations provide a complete census of the \hi\ in and around this nearby Seyfert galaxy, providing missing information on the cold gas flows fuelling the star formation and nuclear activity.
\end{abstract}


\section{Introduction}
\label{intro}
Galaxies are the building blocks of the Universe; their formation and evolution involve a complex interplay of multiple physical processes. The current understanding of galaxy formation suggest that galaxies form from the collapse of primordial gas clouds, which then undergo hierarchical merging and accretion, as well as inflows and outflows, \citep[see][and references therein]{press1974formation, peroux2020cosmic,Watkins2023}. The processes influencing galaxy evolution can be broadly divided into external environmental processes: mergers, tidal interactions, and hydrodynamical forces such as ram pressure stripping; internal processes: including feedback generated by star formation (SF) and active galactic nuclei (AGN), e.g. \citet{hopkins2016stellar}, and slow secular processes such as the redistribution of galactic mass and energy induced by the formation of bars, spiral arms and galactic winds within the disk \citep{kormendy2004secular}. These processes lead to gas entering and being removed from galaxies at different rates and temperatures, thus regulating star formation and the accretion of the supermassive black hole (SMBH). 
This recycling of gas between the galaxy and the environment ultimately regulates the evolution of galaxies \citep[][]{keres2005, Montero2019simba_flows}.

Understanding how AGN and galactic disks are fuelled is one of the main open questions of galaxy evolution. One way the AGN of a galaxy can be sustained is by halo gas accreting onto the galactic disk. 
Accretion of gas into galaxies has been traced by the kinematics of the neutral atomic (\hi) disk of galaxies \citep{sancisi2008accretion}, while accretion of gas onto AGN is difficult to trace directly. In the brightest cluster galaxies, molecular clouds and filaments form stars and are believed to feed the SMBH at the centres of these galaxies \citep{Harrison2018agn_feedback}. In Seyfert galaxies, which are typically hosted by star-forming galaxies, \hi\ disks can trace recent accretion \citep{best2012}. We aim to determine whether this accretion can be linked to the triggering of nuclear activity, and if so, which mechanisms are responsible for it. 
The mechanisms driving accretion involve both external processes, such as gas inflows from the environment, and internal processes such as the redistribution of gas within the galactic disk. These accretion processes can also be influenced by other factors, including galaxy mergers, which can enhance the inflow of gas towards the central AGN.

Galaxy mergers are one of the most common mechanisms that trigger AGN and account for some of the most luminous AGN \citep[][ and references therein]{hopkins2005mergers,Poggianti2017rampress}. 
When two galaxies merge the rotating gaseous disks of both galaxies are disrupted and mixed, then re-collapse, activating star formation and funnelling gas towards the nuclei, thereby fuelling the AGN \citep{hopkins2005mergers, Poggianti2023}. 

Hydrodynamical interactions such as ram pressure have been connected to gas removal and increased local star formation in the outskirts of galaxies \citep[e.g.][]{ellison2018, Poggianti2017rampress,radovich2019gasp}. This is usually associated with galaxies in clusters or massive groups where the dense intergalactic medium efficiently acts as a drag force on their gaseous disks. As a galaxy moves through these dense environments, 
the gas in the disk compresses, driving it towards the central SMBH and potentially triggering nuclear activity \citep{Poggianti2017rampress}.
Meanwhile, less dense gas, such as \hi, is stripped from the galaxy. This stripped gas, having lost its angular momentum, may be pulled back towards the AGN by the galaxy's gravitational potential energy, overcoming the reduced kinetic energy of the gas and possibly triggering the AGN \citep{Poggianti2017rampress}.

Investigating the fuelling of AGN has traditionally been limited to \hi\ absorption studies due to sensitivity limitations of older radio telescopes. A prime example of accretion is seen in NGC\,315, where \citet{Morganti2009ngc315} used \hi\ absorption kinematics to confirm gas accretion onto the galaxy. They found that the immediate environment of NGC\,315 is gas-rich, with one of the absorption features most likely associated with a gas cloud falling into NGC\,315. Another example is the study of \hi\ in PKS\,B1718–649 by \citet{maccagni2014triggers} which identified \hi\ absorption features associated with two small clouds not regularly rotating with the galaxy. \citet{maccagni2014triggers} concluded that the AGN in PKS\,B1718–649 is triggered by local mechanisms, such as accretion of small gas clouds, rather than gas-rich mergers.

In radiatively efficient AGN, accretion onto the SMBH creates a stable accretion disk, and the host galaxies are often associated with star formation (SF) activity \citep{best2012}. These galaxies, often referred to as Seyfert galaxies, typically emit most of their energy through radiative winds and they often host low-power jets. These jets may have an effect on the surrounding ionised gas \citep[e.g. ][]{mingozzi2019magnum, venturi2021magnum}, as well as the molecular and \hi\ gas. IC~5063 is a prime example where the radio jets expanding within the innermost kpc generate a multi-phase outflow extending $\sim1$\,kpc \citep{morganti2015ic5063}. These outflows are detected in molecular gas, \hi, and ionised gas \citep[see][ and references therein]{morganti1998,oosterloo2000,morganti2013,Dasyra2016ic5063}. This removal of gas may eventually deplete the galaxy of its gas supply and halt star formation; a process referred to as negative feedback.

AGN feedback is not limited to the negative feedback effect on star formation but can also have a positive feedback effect, enhancing star formation.
Centaurus~A (NGC~5128) is an example where star-forming regions are triggered by the expansion of the radio jets at the outskirts of its stellar body. The radio jets eject but also compress the multi-phase gas at the edges of the disk and generate a region of recent star formation \citep{santoro2016centA, salome2017centA}. 
Centaurus~A and its outflows have been extensively studied due to the proximity of NGC~5128 (3~Mpc), however, a comprehensive understanding of AGN and stellar feedback and their effects on the evolution of galaxies requires a detailed analysis of a larger sample of nearby AGN and star-forming galaxies. 
We need both sensitive and high-resolution observations to probe the diffuse environment as well as the small-scale structures of a galaxy.

From these studies it is evident that a key component in understanding the feeding and feedback in AGN is the neutral atomic hydrogen (\hi) spatial distribution and kinematics.
\hi\ gas is fundamental for understanding the processes that drive galaxy formation and evolution as it is a prime ingredient of star formation. For the star formation process, \hi\ cools and transforms into molecular gas, \hii, collapsing into a denser state \citep{sancisi2008accretion}. \hi\ has been observed to be involved in fuelling AGN through emission and absorption lines, for example, in NGC\,315 \citep{Morganti2009ngc315} and NGC\,3100 \citep{Maccagni2023ngc3100}.

New radio telescopes, such as the South African MeerKAT telescope, offer a larger field of view, improved spatial and spectral sensitivity and are capable of detecting the \hi\ emission line. This
allows us to greatly improve our knowledge of \hi\ in nearby AGN with short observing times \citep[for example][]{maccagni2021fornaxcold}. MAGNHIFFIC~\footnote{https://magnhiffic.astron.nl} (MeerKAT AGN \hi\ Feeding \& Feedback Investigation Close-by) is an ongoing study of the processes of feeding and feedback in nearby AGN ($D<80$ Mpc) with different energetic outputs, ages, hosts and environments, which leverages sensitive MeerKAT observations of the cold \hi\ to simultaneously probe, for the first time, the small scales near the SMBH, the larger scales of the galactic disks and the environments of these AGN. \ngc\ is one of the first galaxies observed in this project.

\begin{table}
\caption{Basic properties of \ngc.}
\centering
\begin{tabular}{||l c c  ||} 
 \hline\hline
 Parameter & Value & References \\ 
 \hline
  AGN type& Seyfert\,2 & \citet{morris1985velocity} \\
  Centre $\alpha$ (J2000) & 14\textsuperscript{h}32\textsuperscript{m}40.56\textsuperscript{s} & \citet{stuber2023gas} \\
  Centre $\delta$ (J2000) & -44\degr10\arcmin28.56\arcsec\ & \citet{stuber2023gas} \\
  Inclination (\degr) & $27$\degr\ $\pm\,5$\degr\ & \citet{morris1985velocity} \\
  log(SFR [\msun yr$^{-1}$])  &  0.39  & \citet{pan2022gas} \\
  $\text{log}_{10}M_{\star}$ (\msun) & 10.34 & \citet{leroy2021phangs} \\
  Optical redshift & 0.004 & \citet{cresci2015magnum} \\
  Flux [1.4\,GHz] (mJy/beam) & 17 & This work \\
  Average \hi\ mass (\msun)  & $4.4 \times10^{9}$ & This work \\
  \hline
\end{tabular}
\label{properties}
\end{table}

\ngc\ is a Seyfert galaxy where positive feedback has been identified. It is a nearby ($D_{L}$ = 17.3~Mpc, 1"$\approx 83$~pc) barred spiral galaxy, viewed almost face-on (incl = $\sim 27^{\circ} \pm 5^{\circ}$), with the bar oriented in an east-west direction \citep{cresci2015magnum}. There is a clear dust lane parallel to the southern leading edge towards the east of the bar. The basic properties of \ngc\ are given in Table~\ref{properties}. \ngc\ appears to be isolated, with only one other galaxy, PGC\,538542, that is known to be in its environment \citep{kourkchi2017groups}. The redshift of PGC\,538542 is $0.0036\pm1.5\times10^{-4}$, which is similar to \ngc\ \citep{heath2009pgc}. 

\ngc\ has been extensively studied across the electromagnetic spectrum.
In the nucleus of \ngc, there is strong extended line emission from ionised gas which is traced by \oth cones extending $\sim 1.6$~kpc along the bar, observed with MUSE \citep{cresci2015magnum}. 
These double-sided ionisation \oth cones show that the central region has out-flowing ionised gas, which is blue-shifted with a projected velocity out to $\sim450$\kms. 
These outflows point towards star-forming clumps 5" and 10" east of the nucleus, which \citet{cresci2015magnum} suggest is due to positive feedback from these outflows,
implying that positive feedback may be an important mechanism in this galaxy.

The radio structure of \ngc\ was observed with the Very Large Array (VLA) and reported by \citet{morris1985velocity} at 1.5\,GHz, who note the radio emission is aligned with the bar of the galaxy and extends over $\sim1.6$\,\arcmin. \citet{leipski2006radio} also report on VLA radio continuum images, but with greater sensitivity, and find that \ngc\ has two weak, uncollimated, radio jets on either side of the nucleus, which are $\sim30$\arcsec\ long with radio luminosity, $vL_{v}\,[8.4\,\text{GHz}]=0.55\times 10^{13}$~W \citep{leipski2006radio}. 

ALMA observations of \ngc\ show that the molecular CO~$(2-1)$ gas disk extends a few hundred parsecs in size surrounding the AGN and is oriented in a north-south direction \citep{alonso2018resolving}. 
The kinematics of the rotating disk appears to be at a different position angle (PA) and inclination compared to the large-scale disk. The estimated total molecular gas mass of the nuclear disk and the AGN is M$_\text{\hii}\,=\,1.1 \times 10^7$~\msun. \citet{alonso2018resolving} also questions a region of cleared gas in the east which may be associated with a sub-kpc jet, however, there is insufficient evidence to support this assumption.

\ngc\ is part of the large Physics at High Angular resolution in Nearby GalaxieS survey \citep[PHANGS,][]{leroy2021phangs}. The PHANGS survey observed a sample of 90 of the nearest, most massive and star-forming galaxies accessible with ALMA, forming the PHANGSS-ALMA survey which \citet{pan2022gas} reports on. 
\ngc\ is one of these galaxies, with a significant offset (0.4\,dex) from the star-forming main sequence and shows relatively high ($\gtrsim$50 per cent) CO fractions in the ALMA field of view \citep{pan2022gas}. They find that the CO follows the stellar galactic structures well, which, in the case of \ngc, is a barred spiral, suggesting a dynamic origin of these fractions.

Even though \ngc\ has been extensively studied at multiple wavelengths, the role of \hi\ in this galaxy is poorly constrained. Only the \hi\ Parkes All-Sky Survey \citep[HIPASS;][]{barnes2001hipass, meyer2004hipass_cat} integrated \hi\ flux ($\text{S}_{\text{int}}\,=\,56.5\,\pm\,4.7\,\text{Jy\kms}$) is available, and due to the HIPASS resolution of 15\arcmin, the source is unresolved, preventing detailed studies on the circum-nuclear regions \citep[see ][]{Koribalski2004BGC, meyer2004hipass_cat}. 
To address this limitation, we conducted 21-cm MeerKAT observations of \ngc\ that span from the circum-nuclear region near the AGN to the outer regions of the galaxy. These observations aim to provide insight into the mechanisms fuelling the AGN and the broader feedback process in \ngc.

In this paper we present deep \hi\ MeerKAT observation in combination with deep optical observations which, for the first time characterise the \hi\ distribution in this nearby Seyfert galaxy as well as its satellites. In Section~\ref{obs} we describe our observations, calibration, source finding and the ancillary data used in this paper. In Section~\ref{results} we present our results on \hi\ in and around \ngc. Section~\ref{disc_ana} discusses the physical interpretation of our results and compares them to similar findings. The conclusions are presented in Section~\ref{conclusion}.
Throughout this paper we assume $\Lambda$CDM cosmology with cosmological constants values of $H_{0}\,=\,70$\,\kms\,Mpc$^{-1}$, $\Omega_{\Lambda}\,=\,0.7$ and $\Omega_{\text{M}}\,=\,0.3$. At the distance of \ngc, 1\arcsec\ is 83\,pc in the image space.

\section{Observations and data reduction}
\label{obs}
In this section, we describe the observations, data and data reduction techniques used in this study. We utilise MeerKAT L-band observations to image and analyse the \hi\ in \ngc, along with deep optical observations from the VLT Survey Telescope (VST) to identify any optical counterpart to the \hi\ features.

\subsection{MeerKAT L-Band observations } 

\begin{table}
\caption{Neutral hydrogen (\hi) data cube parameters for three cubes at different angular resolutions, covering a $1.5~\text{deg}^2$ field of view and a velocity range of approximately $2000$\,\kms. The table includes the restoring beam, average channel noise (rms) for a channel width of 20\,\kms, column density sensitivity ($3\sigma$ over four 5.5\,\kms channels) and pixel size in arcseconds. Also listed are the cleaning parameters used with \texttt{wsclean}, including the Briggs robust weighting value and whether \textit{uv}-tapering was applied.} 
\centering
\begin{tabular}{||l c c c ||}  
 \hline\hline
  Label&  96\arcsec& 30\arcsec\ &8\arcsec\  \\ 
 \hline
 Restoring beam (\arcsec) &  $97\,\times\,95$& $35\,\times\,25$ & $9\,\times\,6$ \\
 Channel rms ( mJy/beam) & 0.37 &0.18  & 0.25\\
  $N\text{(\hi)}_{3\sigma,20\,\text{\kms}}$ ($\times 10^{19}$ $\text{cm}^{-2}$) & 0.310 &1.42 & 29.5\\  
  Pixel size (arcsec) & 30 & 7 & 2 \\
  Robust weighting & 1.0 & 1.5 & 0.0 \\
  \textit{uv}-tapering & yes & no & no \\
 \hline
\end{tabular}
\label{cube_param}
\end{table}

\begin{figure}
\begin{center}
    \includegraphics[width=\columnwidth]{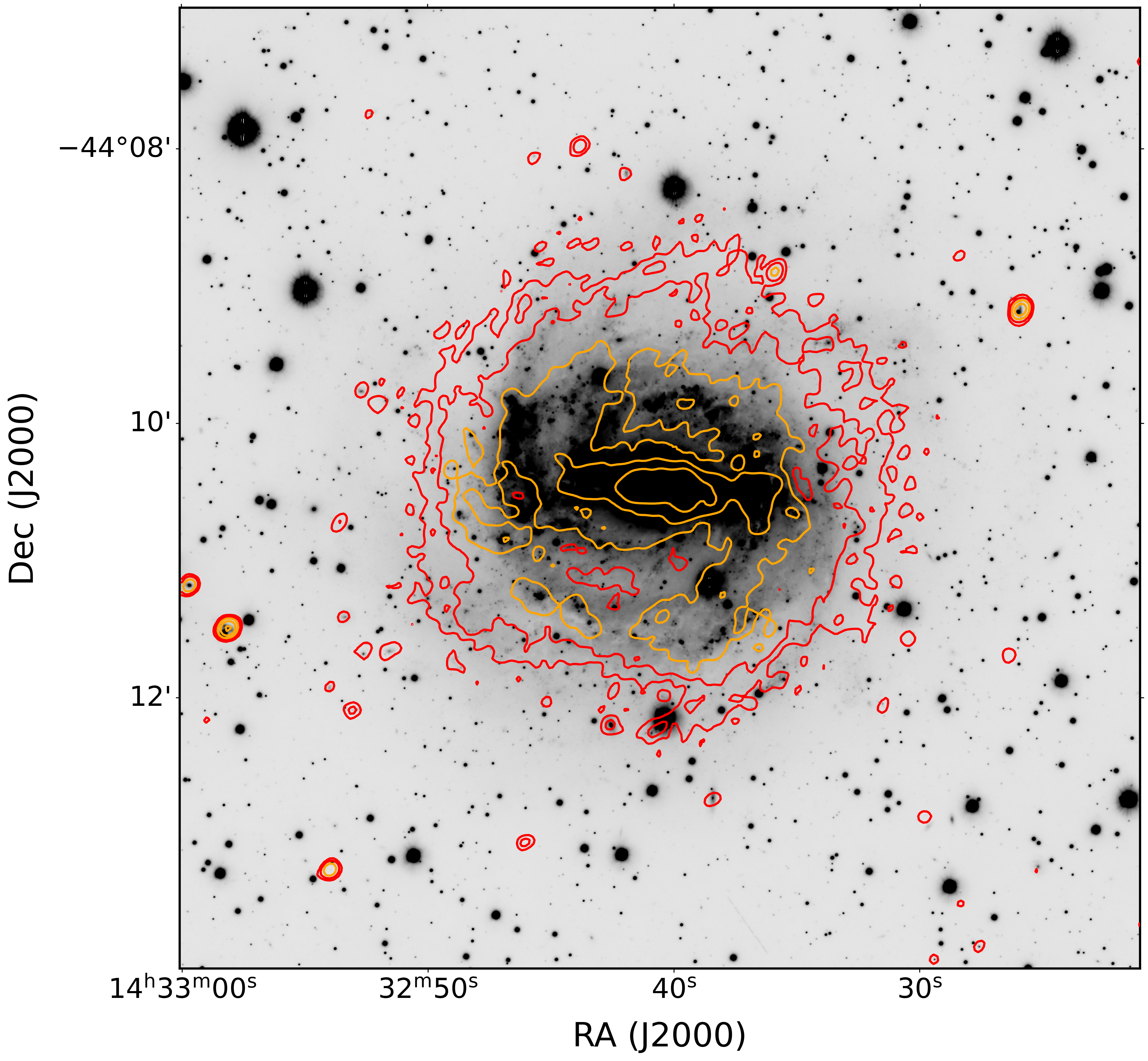}
    \caption[\ngc\ continuum image]{VST r-band optical image of \ngc\ with MeerKAT continuum image contours at $2^{n}\sigma$ where $n\,=2,\,3,\,4,\,5,\,6$ and \mbox{$\sigma\,=\,1.32\times10^{-5}$ Jy/beam}. The inner contours (where $n\,=4,\,5,\,6\,$) are shown in orange to highlight the bar and spiral arm features of \ngc\ visible in the continuum data. The restoring beam of the MeerKAT continuum image is $6.1\text{\arcsec}\times5.0\text{\arcsec}$ with a major axis position angle of -42\degr. The spatial scale is 1.5\arcsec per pixel. }
    \label{cont}
\end{center} 
\end{figure}

The MeerKAT observations of \ngc\ were conducted at night on 2021-03-19 and 2021-06-05 (SCI-20210212-FM-01, PI Maccagni) in two separate 5.5-hours tracks. The target was observed while rising and setting to maximise coverage of the uv-plane. Here, we present the observations in the frequency range $1340.0-1444.5$\,MHz at the original spectral resolution of (26.12 kHz).

The observations were processed with {\tt CARACal}~\citep[][]{jozsa2020}, a Python-based, containerised pipeline commonly used for the reduction of MeerKAT continuum and \hi\ observations~\citep[e.g.][]{maccagni2020,namumba2021,Ianjamasimanana2022,serra2023,deblok2024}. The data reduction strategy follows that of the MeerKAT Large Survey programmes, MeerKAT Fornax Survey~\citep{serra2023} and MHONGOOSE \citep[][]{deblok2024}, to which we refer for a detailed description. Here, we briefly summarise the three main stages of the process: cross-calibration, self-calibration and spectral line imaging.

In the first stage, we calibrated the single 5.5-hour tracks independently. After cross-calibration of the target, the single tracks were flagged for radio frequency interference through automated AOflagger routines \citep{offringa2012} and self-calibrated using {\tt wsclean} \citep[][]{offringa2014wsclean} and {\tt cubical} \citep[][]{Kenyon2018}. Self-calibration was performed over the full 100\,MHz band divided in four spectral bins. The self-calibration solutions were then transferred to the
measurement sets at the original spectral resolution using {\tt crystalball}. The continuum model was then subtracted from the measurement sets, and additional continuum subtraction was performed by fitting and subtracting a first-order polynomial to each point in the \textit{uv}-plane independently (excluding the spectral range of known \hi\ emission). A last flagging routine was then run to remove the residual broadband RFI typical of the short baselines of interferometers near $u=0$~\citep[][]{hess2015,carignan2016,heald2016,maccagni2020}. 

In the second stage, we jointly deconvolved and self-calibrated the two tracks together to produce the most sensitive 100\,MHz continuum image of the galaxy and the 2000~\kms-wide multi-scale \hi\ data cubes.

The final stage involved combining the continuum-subtracted measurement set to generate the data cubes, with a 5.5~\kms\ channel width and at multiple resolutions. These resolutions were produced using different Briggs robust weighting values and, in some cases, applying \textit{uv}-tapering, see Table~\ref{cube_param}. 
For the purpose of this paper, we focus on three spectral cubes with resolutions of 96\arcsec, 30\arcsec\ and 8\arcsec, which, at the distance of \ngc, correspond to approximately $8.0\,\text{kpc}, 2.5\,\text{kpc}, 0.7\,\text{kpc}$, respectively. These MeerKAT observations are sensitive to column densities of $\sim\,10^{18}, \sim\,10^{19}, \text{and} \sim\,10^{20} \,\text{cm}^{-2}$ for the 96\arcsec, 30\arcsec\ and 8\arcsec\ cubes, respectively. Table~\ref{cube_param} presents the noise levels in each cube with a channel width of 5.5\,\kms as well as the column density sensitivity, defined at the $3\sigma$ level over four channels ($\sim20$\,\kms), which marks the sensitivity limit of MeerKAT in these observations.

We created these multi-resolution cubes to investigate the different features they reveal. The 96\arcsec\ cube enables us to examine the diffuse gas in the galaxy, while the 8\arcsec\ cube provides the high resolution required to study the high column density \hi\ in the star-forming disk and \hi\ absorption. The 30\arcsec\ cube offers a good balance between capturing diffuse gas and resolving high-resolution features and is therefore used for most of the analysis in this paper.

The 1.4\,GHz MeerKAT continuum image of \ngc\ is shown in Figure~\ref{cont}, overlaid in red and orange contours on the optical VST image. The continuum image has a bandwidth of 100~MHz and an rms noise level of $0.13\,\umu$Jy. At a resolution of 5.5\arcsec\ ($\sim0.5$~kpc), the radio continuum emission traces the star-forming disk, central bar and the spiral outer arms. The orange contours in Figure~\ref{cont} emphasise the central bar and spiral arms in \ngc.

\subsection{Source Finding}
\label{sofia}

\begin{figure*}
    \includegraphics[width=0.8\textwidth]{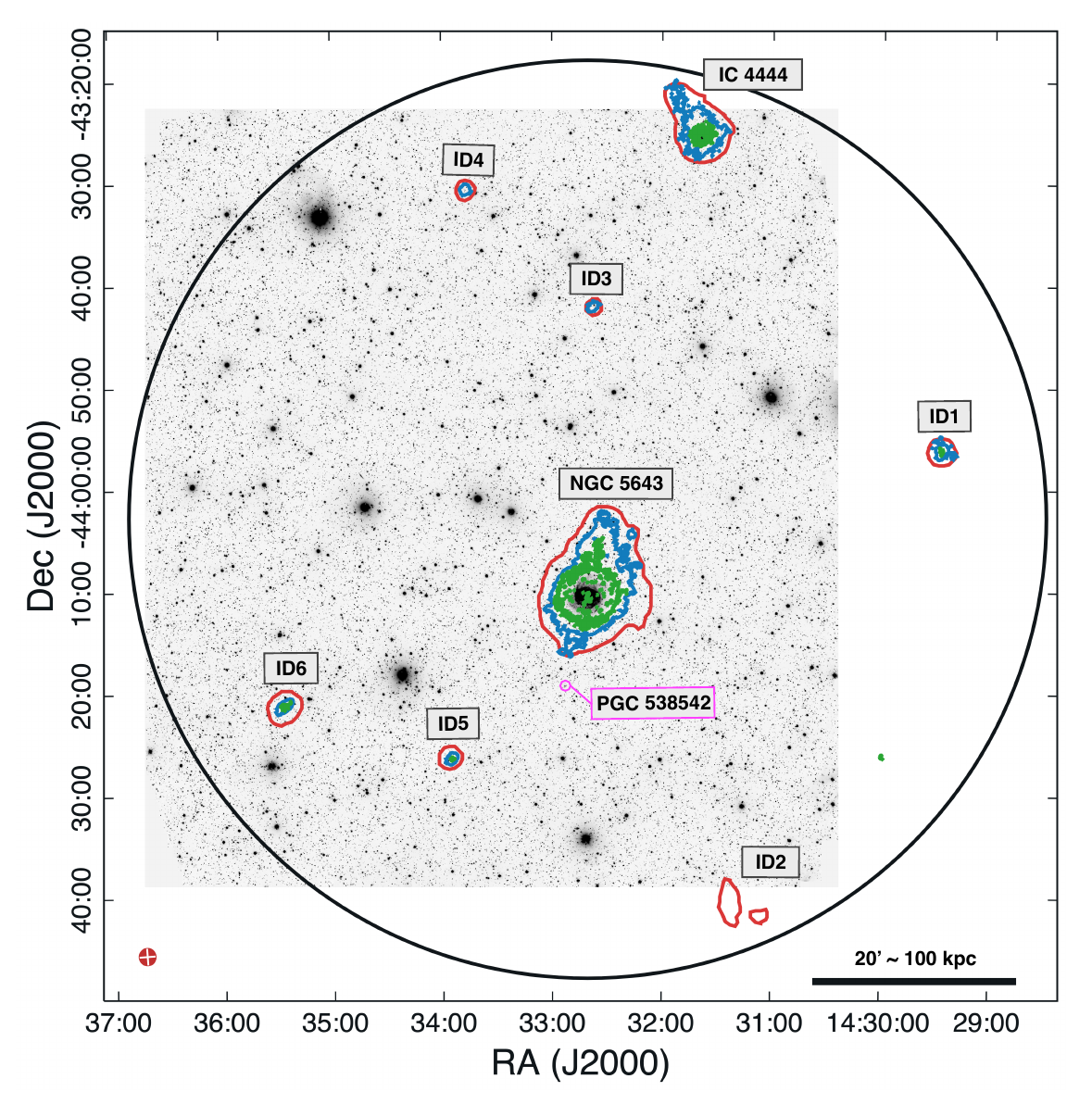}
    \caption[VST with MeerKAT column density]{Optical g-band VST image of \ngc\ with \hi\ column density contours ($3\sigma$ over four channels) at [$3.1\times10^{18}, 1.4\times10^{19}, 2.9\times10^{20}$]~$\text{cm}^{-2}$ for each cube with resolution 96\arcsec\ (red), 30\arcsec\ (blue) and 8\arcsec\ (green), respectively. The surrounding sources (excluding IC\,4444) are detected for the first time with these \hi\ observations. 
    The closest \hi\ cloud to \ngc\ is ID5, which is $\sim100$\,kpc away. The total image area of the MeerKAT observation is enclosed in a 1.5\,degree diameter black circle. The beam for the 96\arcsec\ cube is shown as a red ellipses in the bottom left. The only previously known galaxy in the environment of \ngc\ is PGC\,538542, which is highlighted in magenta. 
    }
    \label{full_fov}
\end{figure*}

\begin{table*}
\caption{Information on the \hi\ sources within the 1.5\,deg$^2$ field of view from the {\tt SoFiA-2} catalogue and our calculations. In the first column, the newly detected sources are named according to the MeerKAT naming system. The second column provides the shorthand notation for the new sources, as used throughout the paper. The table includes the coordinates of each source, along with the separation, which represents the spatial distance of the sources from \ngc. The velocity difference ($\Delta v$) is given with respect to \ngc, based on the radial velocity $v_{\text{rad}}$. The full width at half maximum (FWHM) for each source is listed, including the associated channel width error. The integrated flux ($\text{S}_{\text{int}}$) was used to calculate the \hi\ masses ( M$_{\text{\hi}}$ ) from the 30\arcsec\ data, with an estimated 10\,per cent error.  }
\fontsize{7}{8}
\selectfont{
\begin{tabular}{ccccccccccc}
\hline
\hline
Name                     &id      & RA$_{\text{J2000}}$   & DEC$_{\text{J2000}}$      & separation & separation & $v_{\text{rad}}$  & $\Delta v$ & FWHM& $\text{S}_{\text{int}}$ & M$_{\text{\hi}}$ \\
& & (\,$^h~^m~^s$\,) & (\,$^d~^m~^s$\,)       & (\,\arcmin\,) & (\,kpc\,)& (\,\kms\,) & (\,\kms\,)         & (\,\kms\,) & (\,Jy\,\kms\,) & (\,$\times 10^{8} \mathrm{M_{\odot}}$\,)    \\
\hline 

MKT\,J142928.8-435624    & ID1         & 14 29 28.8  & -43 56 24    & 38& 189 & 1137 & -55    & 39\,$\pm 5.5$ & 0.27 & 0.56\,$\pm$\,0.06 \\
MKT\,J143116.7-444045    & ID2         & 14 31 16.8  & -44 40 48    & 34& 169 & 261 & -931   & 16\,$\pm 5.5$& 0.22  & 0.37\,$\pm$\,0.04 \\
IC\,4444                  &             & 14 31 38.4  & -43 25 12   & 47& 234 & 1950 & 758  & 165\,$\pm 5.5$ & 3.73 & 15.3\,$\pm$\,1.53\\
MKT\,J143238.4-434200    & ID3         & 14 32 38.4  & -43 42 00    & 28& 139 & 1898  & 706     & 35\,$\pm 5.5$& 0.09 & 0.11\,$\pm$\,0.01 \\
\ngc\                    &             & 14 32 40.8  & -44 10 12    & 0 & 0   & 1192  & 0  & 191\,$\pm 5.5$ & 62.3 & 43.9\,$\pm$\,4.39\\
MKT\,J143348.0-433036    & ID4         & 14 33 48.0  & -43 30 37    & 41& 204 & 902  & -290   & 25\,$\pm 5.5$  & 0.10& 0.28\,$\pm$\,0.03 \\
MKT\,J143355.2-442624    & ID5         & 14 33 55.2  & -44 26 24   & 21& 105 & 1923 & 731   & 22\,$\pm 5.5$ & 0.13 & 0.13\,$\pm$\,0.01\\
MKT\,J143526.4-442100    & ID6         & 14 35 26.4  & -44 21 00    & 32& 159 & 822  & -310  & 30\,$\pm 5.5$& 0.35 & 0.53\,$\pm$\,0.05 \\

\hline
\end{tabular}}
\label{table:sources2}
\end{table*}

We manually searched the three data cubes for \hi\ sources and identified seven \hi\ detections within the 1.5 deg$^2$ field of view surrounding \ngc. Among these surrounding sources, \hi\ has only been detected in IC\,4444, with a HIPASS integrated flux of $\text{S}_{int}\,=\,24.9\,\pm3.0$ \citep{Koribalski2004BGC}. To verify the validity of the newly identified sources, we used the automated Source Finding Application \citep[{\tt SoFiA-2};][]{westmeier2021sofia}. {\tt SoFiA-2} constructs a source mask by using a smooth and clip (S+C) algorithm, which works by smoothing the data cube on multiple user-defined scales. We set the source detection threshold to signal-to-noise ratio (SNR) of four. 
This multi-scale smoothing and clipping method is capable of identifying relatively faint and extended features that may have been missed by visual inspection or by applying a single threshold cut-off. 

For the automated source finder, we applied the same {\tt SoFiA-2} parameters across all cubes, with the exception of the reliability parameter, which was set to 80 per cent, 75 per cent and 70 per cent for the 96\arcsec, 30\arcsec\ and 8\arcsec\ cube, respectively. The reliability threshold was lowered to include a few visually identified sources that were not automatically detected but have an optical counterpart (see Section~\ref{optical}).
 
We generated the flux density, velocity field and velocity distribution maps of the \hi\ detected within the field. Figure~\ref{full_fov} presents these detections overlaid on the VST g-band, optical image (described below). We characterised the \hi\ properties of the detected sources and summarised them in Table~\ref{table:sources2}.

\subsection{Optical observations}
\label{optical}
Assessing the presence of low-surface-brightness galaxies or stellar features associated with the \hi\ emission or absorption detected by MeerKAT requires deep optical observations. As part of the MAGNHIFFIC project, we are collecting sensitive optical data for our targets, either from the Dark Energy CAmera Legacy Survey (DR10-DeCALS), where available, or through targeted observations with the VLT Survey Telescope (VST). 

The field of view of the VST-OmegaCAM closely matches that of MeerKAT,  covering approximately $1\,\text{deg}^{2}$, and allows for rapid imaging down to a deep surface brightness limit of \mbox{27\,mag\,arcsec$^{-2}$}. This depth is crucial for identifying optical counterparts to the detected \hi\ and for detecting pristine \hi\ clouds. The combination of deep MeerKAT and VST observations, whether from the Fornax Deep Survey \citep{Venhola2019fornax_deep, iodice2016stellar} or from VEGAS \citep{2015vegas, Spavone2017vegas}, has been instrumental in studying the environment of nearby AGN, such as Fornax A~\citep{kleiner2021meerkat} and NGC\,3100~\citep{Maccagni2023ngc3100}. These datasets have also enabled the discovery of the lowest gas-mass surface brightness galaxy beyond the Local Group~\citep[][]{maccagni2024}. 

\ngc\ was observed with the VST in the g and r band in April 2024 (project P114, PI Maccagni). A zoomed-in r-band image of the target is shown in Figure~\ref{cont}, while the full-field image, used to identify the sources within $\sim$1~deg$^2$ of \ngc, is presented in Figure~\ref{full_fov}.

\subsection{Ancillary data}
\label{anc_data}
\ngc\ has been extensively studied across multiple wavelengths by several surveys. 
It is part of the PHANGS sample and has been observed with ALMA \citep{leroy2021phangs}. The ALMA data are available from Canadian Astronomy Data Center (CADC)\footnote{https://www.canfar.net/storage/list/phangs/RELEASES/PHANGS-ALMA/}.
In this paper, we consider ALMA observations of low-J carbon monoxide, CO(2-1), which trace molecular hydrogen, \hii, throughout the star-forming disk \citep[see][for example]{leroy2021phangs}. Molecular gas in \ngc\ has been detected out to a projected radius of $\sim7$\,kpc, with an angular resolution of 1.3\arcsec\ \citep{leroy2021phangs}.

\section{Results}
\label{results}
In this section, we present the results from our MeerKAT observations, focusing on the \hi\ in \ngc\ and its surrounding environment.

\subsection{Full field of view}
\label{Full_FOV_sec}
We imaged a 1.5~deg$^2$ field of view ($\sim 448$~kpc at D$_L=17$\,Mpc, the distance of \ngc), despite the sensitivity of MeerKAT dropping beyond 1~deg. This larger area was chosen because we detected \hi\ emission from multiple galaxies at the edge of the field of view, allowing for a more complete characterisation of their \hi\ properties (see figure~\ref{full_fov}). 
Only \ngc\ and IC\,4444 have been previously catalogued in \hi, the other detections are new and named using the MeerKAT convention, given in Table~\ref{table:sources2}. 
We assigned numbered IDs to these new detections for brevity in this paper. The contours in Figure~\ref{full_fov} represent the \hi\ column density sensitivity at a $3\sigma$ level, assuming a line width of $\sim 20$\,\kms (four channels) for the 96\arcsec, 30\arcsec\ and 8\arcsec\ data cube. The projected spatial separation of the \hi\ sources from \ngc\ 
ranges from 21\arcmin\ to 41\arcmin\ (104~kpc to 206~kpc), while projected velocity separation ranges from  55~\kms~ to 758~\kms. The neighbouring galaxy, PGC\,538542, as mentioned in Section~\ref{intro}, shows no detectable \hi\ emission. It is likely that the \hi\ in this galaxy has been stripped by the more massive \ngc.

\hi\ gas in \ngc\ is detected in emission (Section~\ref{hi_emi_sec}) and absorption (Section~\ref{hi_abs_sec}). The \hi\ mass of \ngc\ is \mbox{$M_{\matHI}$ $\approx 4.4 \times 10^{9}$\,\msun}, which is an order of magnitude lower than its stellar mass, $\text{log}(\frac{\text{\mstar}}{\text{\msun}})=10.3$ \citep{leroy2021phangs}. Excluding IC\,4444, the surrounding \hi\ sources have an average \hi\ mass of $M_{\matHI}$~$\approx~3~\times~10^{7}$\,\msun.

We use the VST image to determine whether these surrounding sources have optical counterparts. The sensitivity of the VST image reaches a surface brightness of $\sim\,27$\,mag/arcsec$^{2}$, allowing us to detect galaxies with stellar masses of $10^6$\,\msun. For unresolved sources at the distance of \ngc, the brightness limit corresponds to a stellar mass limit of $10^4$\,\msun. Therefore at 12~arcsec$^2$, we obtain a $3\,\sigma$ signal, which constitutes a reliable detection, corresponding to an apparent magnitude of 25.8 and log($\frac{\text{\mstar}}{\text{\msun}})\,=\,4.4$.

When compared to the VST image ID4, ID5 and ID6 have identifiable optical counterparts. The presence of a foreground star makes it difficult to determine whether ID3 has an optical counterpart. ID1 and ID2 fall outside the field of view of the VST image. Further information on these \hi\ sources can be found in the appendix, Figure~\ref{id_plot}.

For the first time, we are able to identify these dwarf galaxies and associate them with the environment of \ngc\ and IC\,4444. These discoveries indicate that \ngc\ is not as isolated as originally thought, which may have implications for its evolution, as explored in this paper.

\subsection{\hi\ emission in \ngc}
\label{hi_emi_sec}

\begin{figure}
  \centering
   \includegraphics[width=0.45\textwidth]{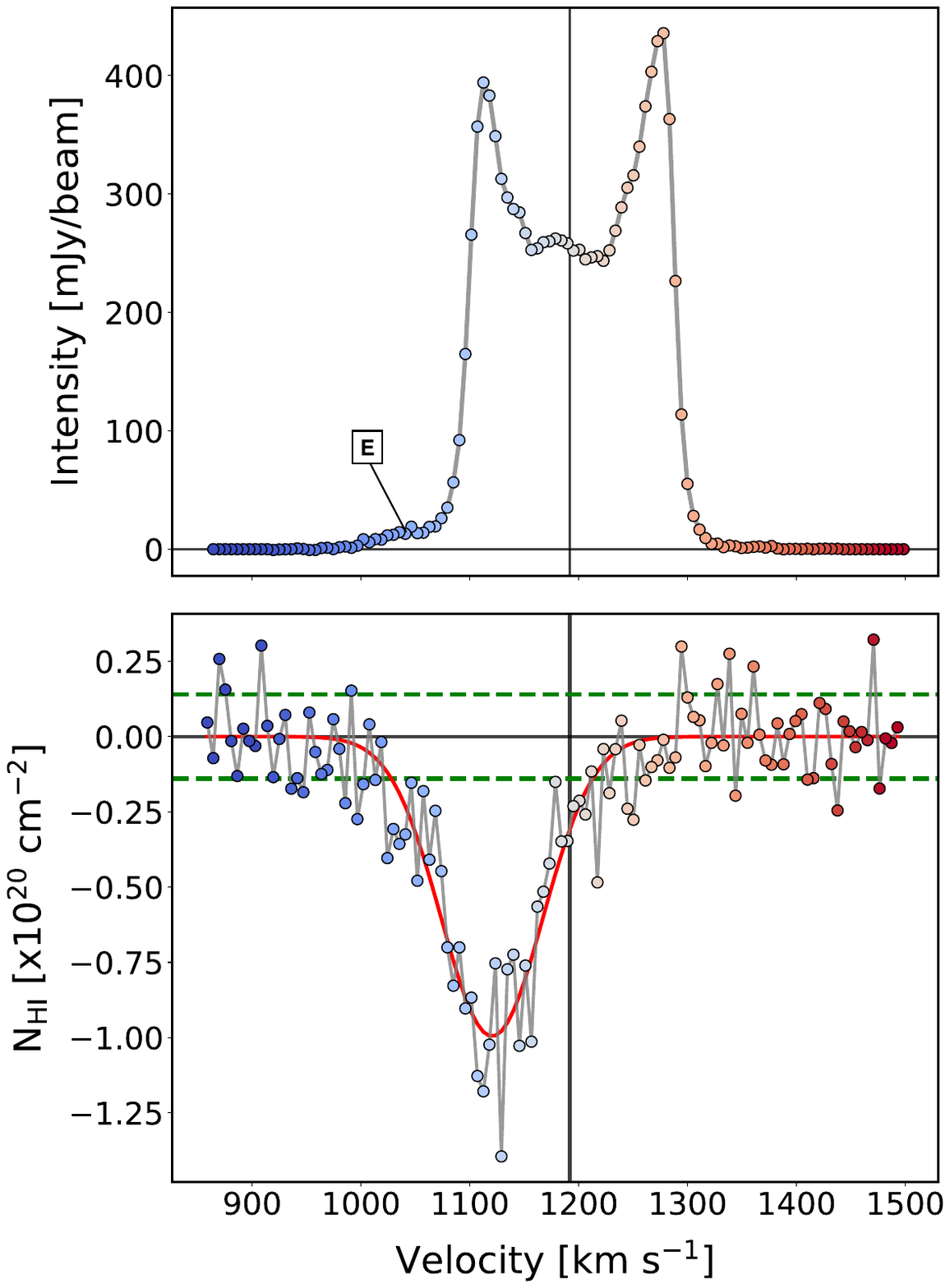}
  
\caption[\hi\ absorption spectrum ]{\textit{Top panel:} \ngc\ \hi\ emission spectrum from the 30\arcsec\ cube. An asymmetry is observed between 1000\,\kms and 1100\,\kms, corresponding the northern \hi\ tail (E). \textit{Bottom panel:} \hi\ absorption column density spectrum towards the centre of \ngc. The absorption spectrum was extracted from the central beam (diameter of 664\,pc) of the 8\arcsec\ data cube. The average rms for the channels in the 8\arcsec\ cube is indicated by the horizontal green lines at $\pm 1.4 \times 10^{19} ~\text{cm}^{-2}$.
In both panels, the black vertical line represents the systemic velocity of the galaxy, $v_{\rm sys}=1192$~\kms. }
\label{abs}
\end{figure}

\begin{figure*}
\begin{center}
    \includegraphics[width=0.85\textwidth]{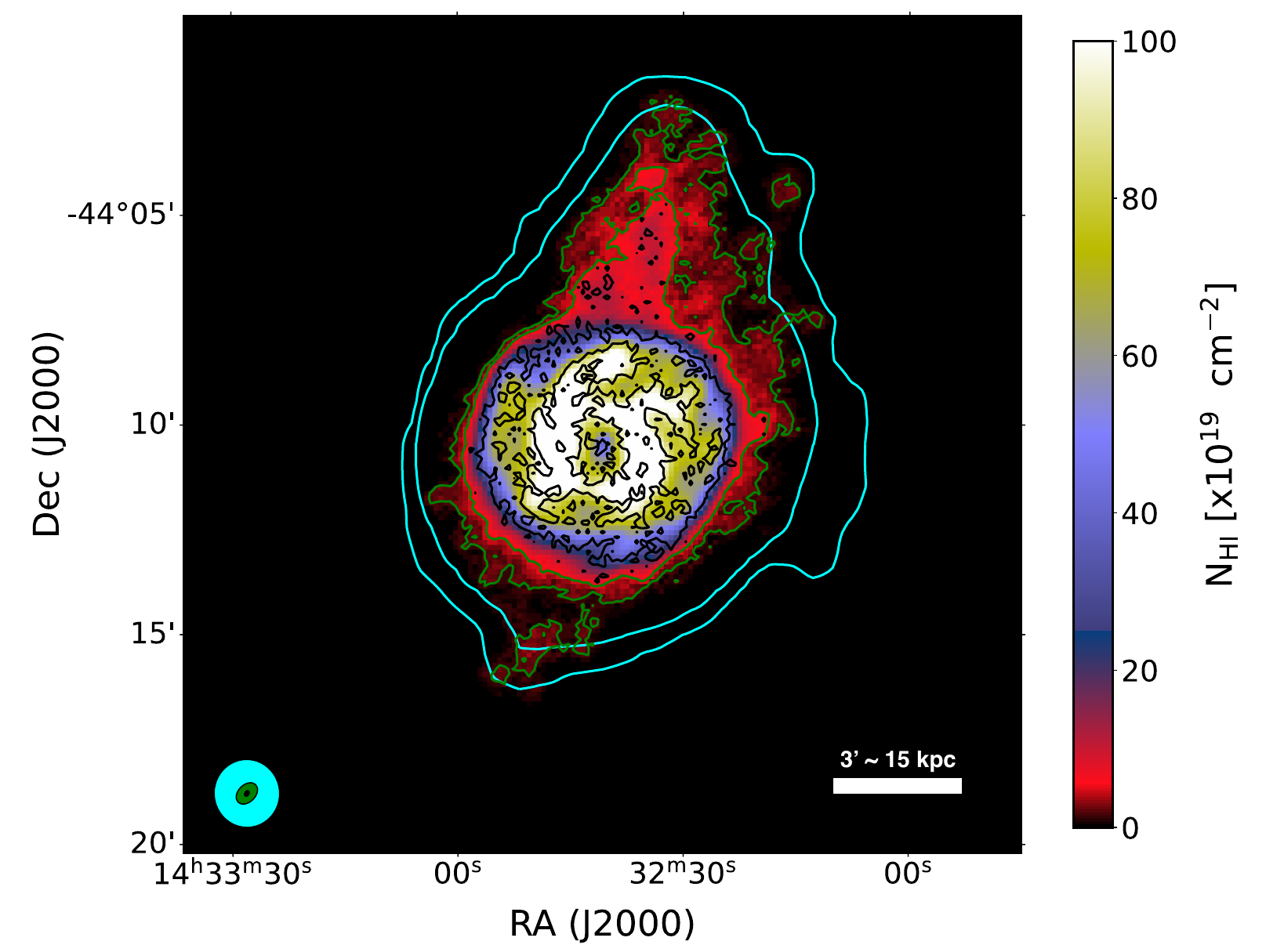}
    \caption[Intensity map]{\ngc\ intensity map for the 30\arcsec\ cube with $3\sigma$ and $9\sigma$ \hi\ column density contours for each cube. The 96\arcsec\ intensity map contour levels are [3.1,~9.3]~$\times10^{18}~\text{cm}^{-2}$ (cyan). The 30\arcsec\ intensity map contour levels are [1.7,~5.2]~$\times10^{19}~\text{cm}^{-2}$ (green). The 8\arcsec\ intensity map contour levels are [2.9,~8.9]~$\times10^{20}~\text{cm}^{-2}$ (black).
    }
    \label{mom0_maps}
\end{center} 
\end{figure*}

Focusing on \ngc, we analyse the \hi\ emission using the 96\arcsec, 30\arcsec\ and 8\arcsec\ data cubes. 
The integrated flux profile of the 30\arcsec\ cube (top panel of Figure~\ref{abs}) traces the \hi\ disk. Additionally, \hi\ absorption is detected in the centre of \ngc\ (bottom panel of Figure~\ref{abs}), which we discuss in the next section. The integrated flux profile exhibits a symmetric double-horn shape, tracing the rotation of the disk; however, a wing is present at blue shifted velocities.The integrated flux of \ngc\ with MeerKAT is $\text{S}_{\text{int}}=62.3$\,Jy/\kms, which suggests \ngc\ is 10 per cent more \hi\ massive than inferred in HIPASS. However a similar result was seen in \citet{deblok2024}, where they note that this discrepancy may be due to some \hi\ emission being subtracted during the data processing, particularly when the bandpass correction method was applied. Another possible explanation is that the flux densities of \hi\ bright sources in the HIPASS data may be affected by the gridding used in the HIPASS pipeline \citep{deblok2024}. With MeerKAT we are able to resolve the disk, as shown in the \hi\ flux density distribution (moment~0) map, Figure~\ref{mom0_maps}.
 
Figure~\ref{mom0_maps} shows the moment~0 map for the 30\arcsec\ cube, with $3\sigma$ and $9\sigma$ column density contours of the 96\arcsec, 30\arcsec\ and 8\arcsec\ intensity map in cyan, green and black, respectively. 
The \hi\ is distributed in a circular disk with a 30\,kpc diameter, highlighted in blue. The high column density gas within the disk traces the spiral arms well, highlighted in green and white.

\begin{figure}
  \centering
   \subfloat{\includegraphics[width=0.5\textwidth]{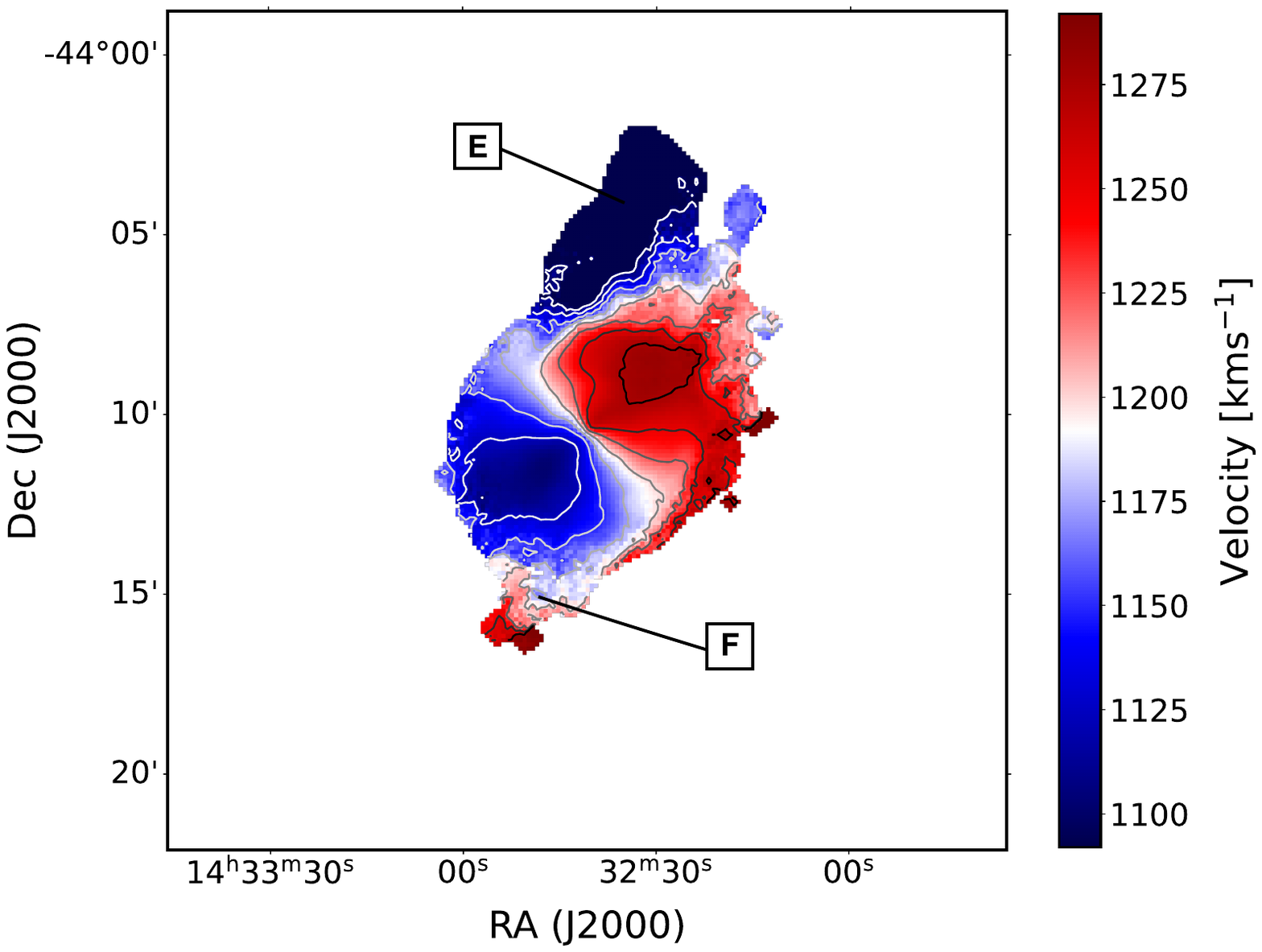}}
  \hfill
   \subfloat{\includegraphics[width=0.5\textwidth, trim=20 0 0 0, clip]{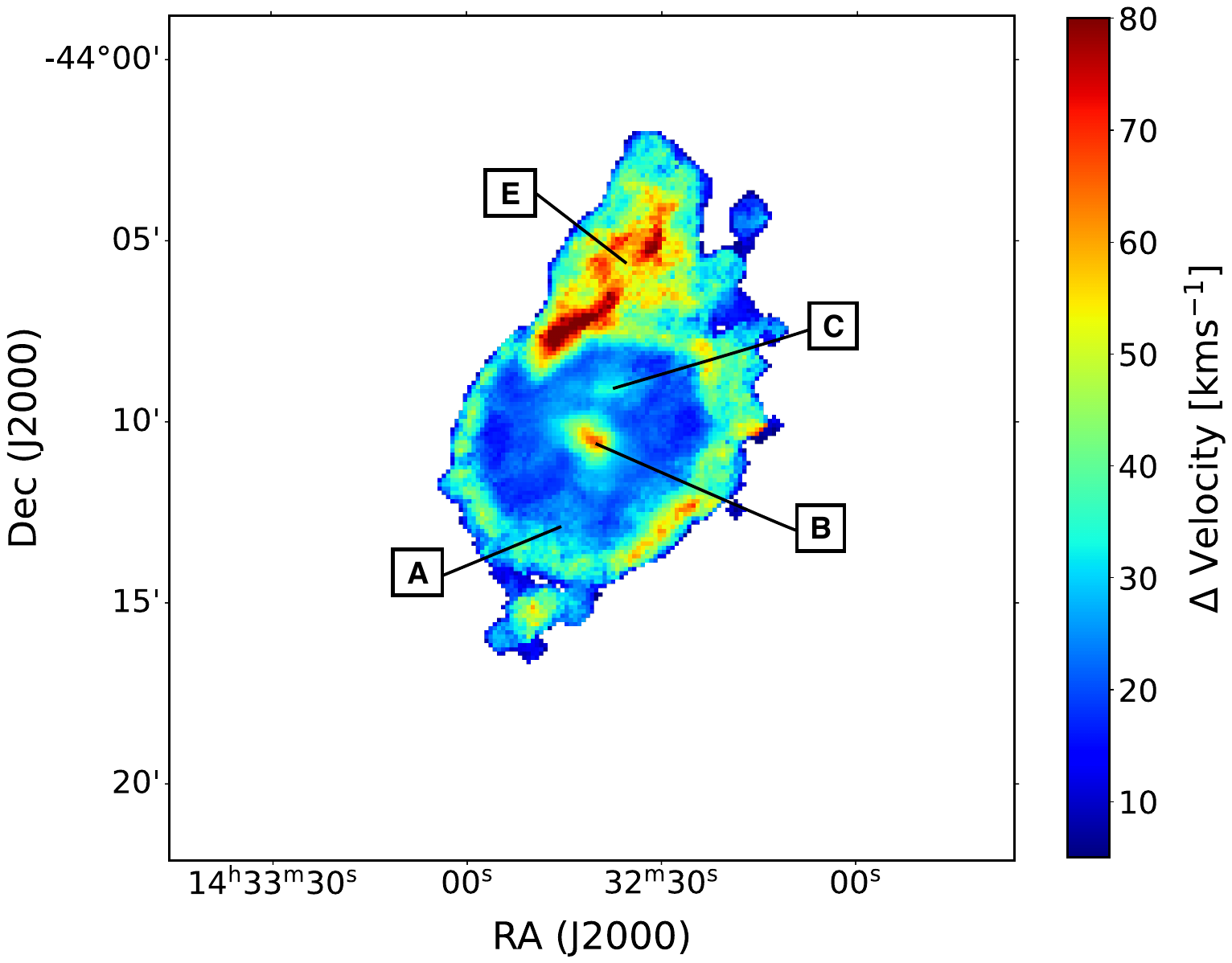}}
\caption[\ngc, velocity maps]{Kinematics of \ngc\ from the 30\arcsec\ cube. \textit{Top}: Velocity map (moment 1) centred on the systemic velocity of \ngc\ (1192\,\kms). The contours are from 1100\,\kms\ to 1275\,\kms, increasing in steps of 25\,\kms. \textit{Bottom}: Velocity dispersion map (moment 2) of \ngc. The annotations highlight features discussed throughout the paper.}
\label{mom1_maps}
\end{figure}

The velocity field of \ngc\ is shown in the top panel of Figure~\ref{mom1_maps} for the 30\arcsec\ cube, revealing a regularly rotating \hi\ disk with position angle, PA\,$\sim\,315^{\circ}$. This is consistent with the PHANGS position angle (PA\,=\,317\degr), measured from the rotation axis of the 13~kpc central molecular disk \citep{lang2020phangs}.
The bottom panel of Figure~\ref{mom1_maps} presents the velocity dispersion map. The average velocity dispersion in the disk is $\sigma_{\text{v}}\,\sim\,20$\,\kms. Around the edge of the galactic disk, particularly in the northern and southern regions, we see higher velocity dispersions ($\geq\,30$\,\kms), likely tracing more turbulent gas. 
This is similar to what is observed in NGC\,2403, another nearby star-forming galaxy, where the velocity dispersion in the disk ranges from 8 to 12\,\kms. In the central region and spiral arms of NGC\,2403, the velocity dispersion reaches 10-15\,\kms, producing a beard effect in the position velocity diagrams, which is linked to ongoing star formation in these regions \citep[see ][]{sancisi2008accretion,Fraternali2017review,deblok2014N2403}. While the velocity dispersion in the anomalous infalling gas filament in NGC\,2403 is higher, ranging from $20$ to $50$\,\kms~ \citep[see][]{fraternali2002n2403, Veronese2023ngc2403}. 

The most peculiar feature of the \hi\ in \ngc\ is the low-column density northern tail, which we detect in all data cubes. This tail extends $\sim 30$~kpc from the nucleus of \ngc. 
Most of the tail has low column densities, ranging from $0.3\,-\,6\,\times 10^{19}\text{cm}^{-2}$, and is most prominently observed in the 96\arcsec\ cube, even though it is resolved with only $\sim2\,$beams. The tail is spatially resolved in the 30\arcsec\ data cube but only faintly visible in the 8\arcsec\ cube. 

\begin{figure*}
\begin{center}
    \includegraphics[width=0.65\textwidth]{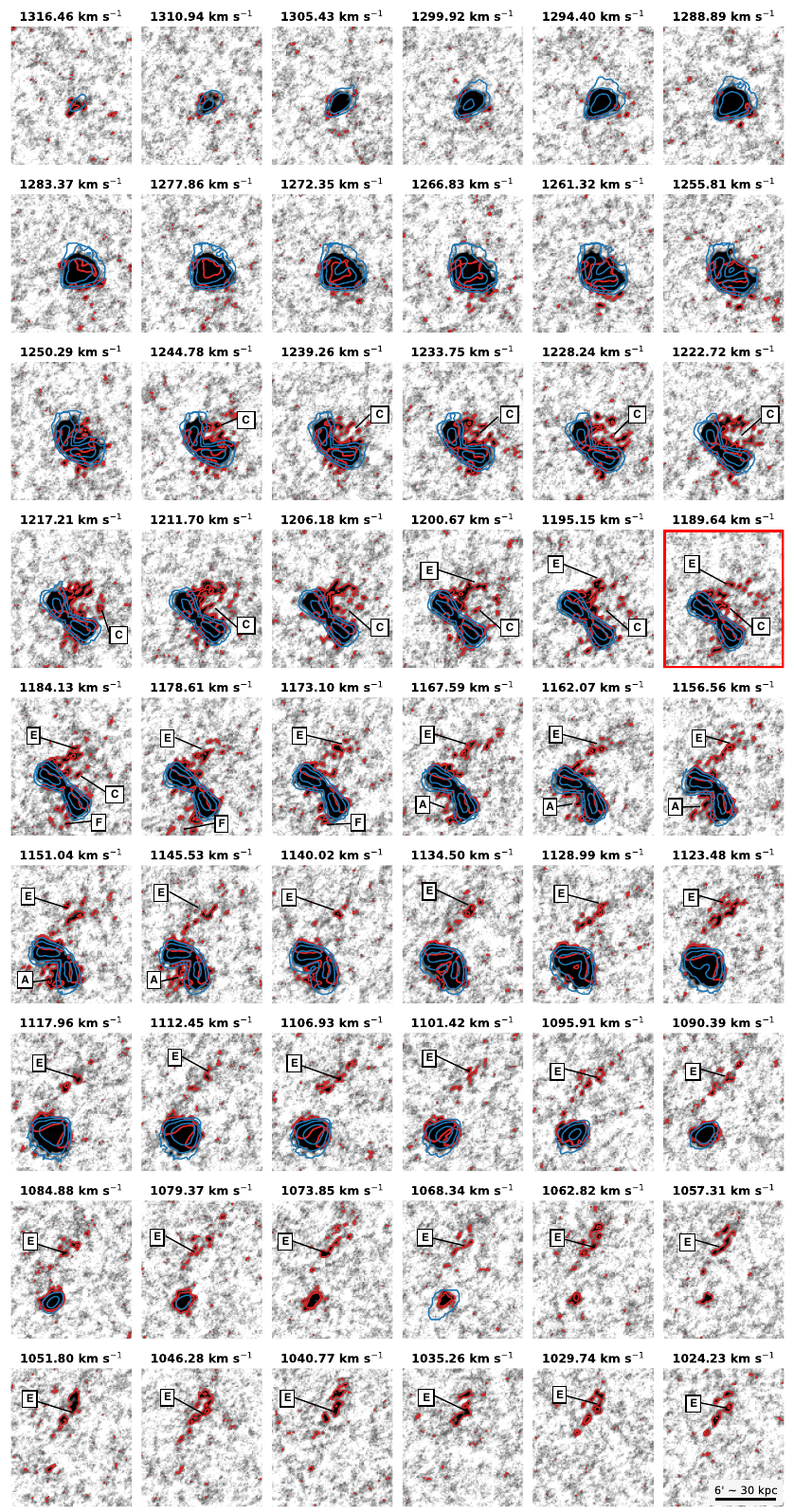} 
    \caption[Channel maps]{Channel maps of 30\arcsec\ cube, with the model (Section~\ref{bbarolo}) in blue contours at [0.1, 1.0, 10]\,mJy/beam. The difference between the data and the model (the residual) is shown in red contours at [0.5, 1.0]\,mJy/beam isolating the tail and other non-rotating gas. The channel closest to the systemic velocity is marked with a red boarder. The labels provide an indication of the features discussed throughout the paper. 
    }
    \label{chan_maps}
\end{center} 
\end{figure*}

\begin{figure}
  \centering
   \subfloat{\includegraphics[width=0.5\textwidth, trim=40 60 40 60, clip]{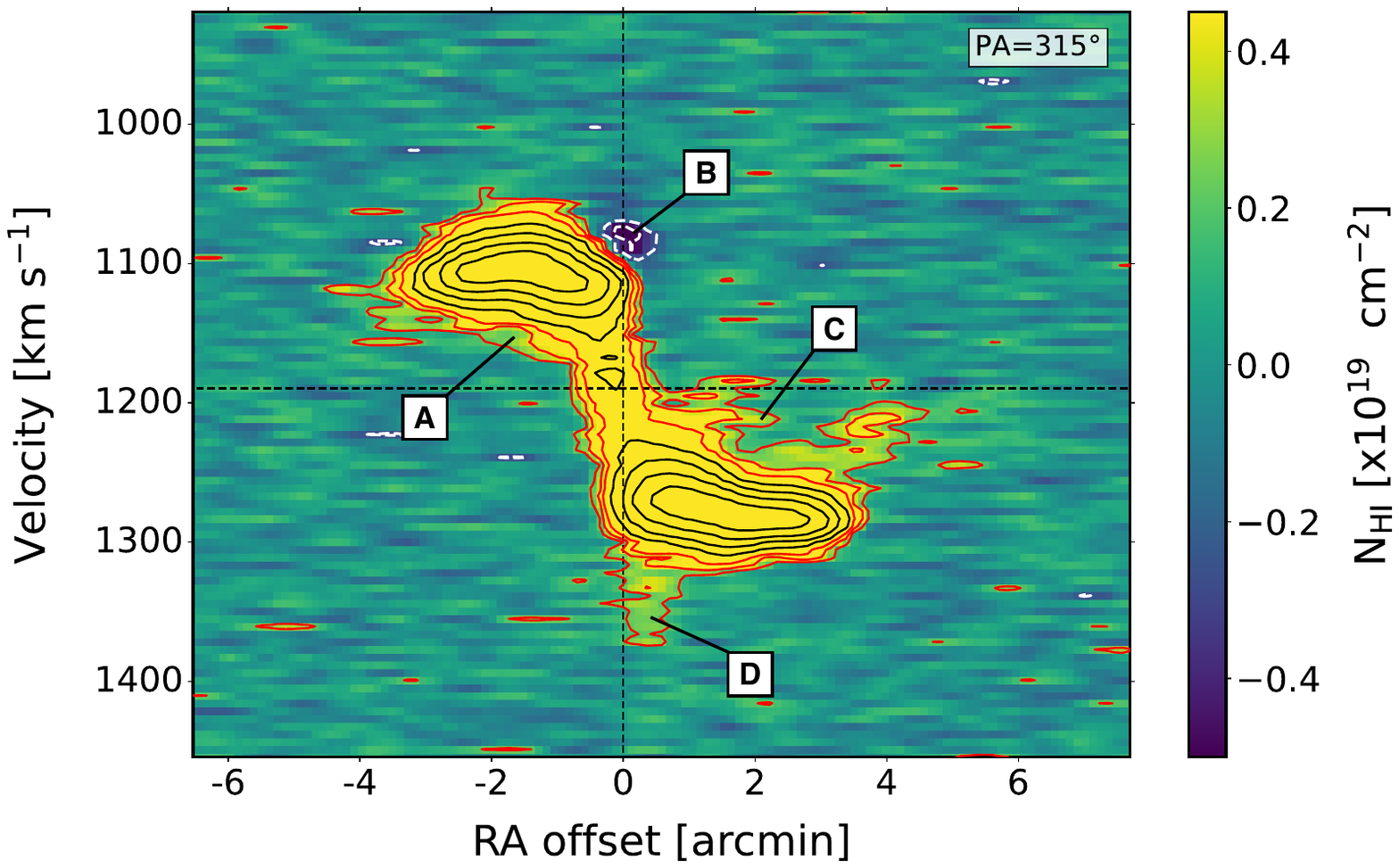}}
  \hfill
   \subfloat{\includegraphics[width=0.5\textwidth, trim=40 60 40 60, clip]{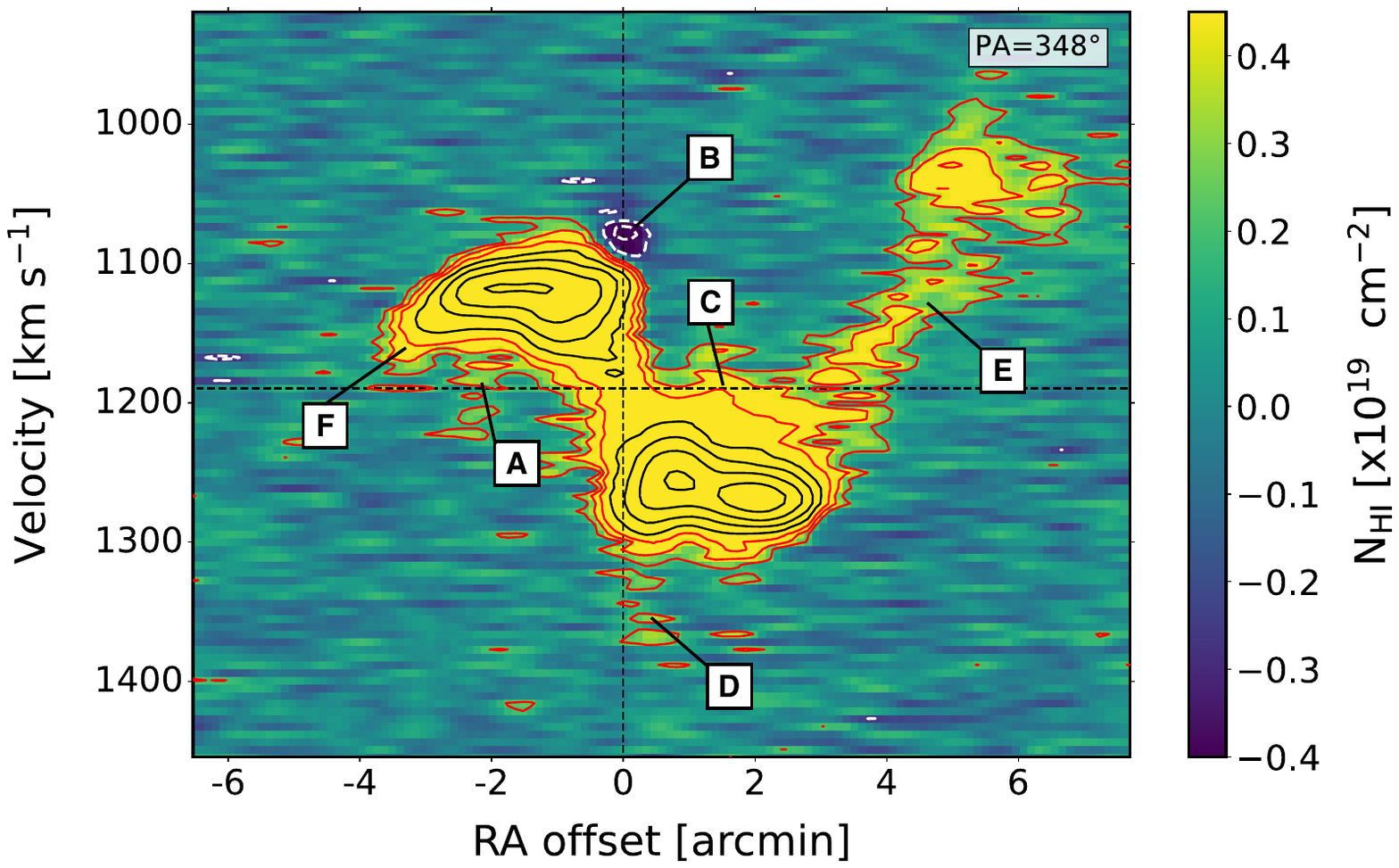}}
\caption[position velocity data only]{\ngc\ position velocity diagrams. \textit{Top panel:} Slice along the major rotation axis, PA$\,=\,315$\degr. \textit{bottom panel:} Slice along the tail feature, PA$\,=\,348$\degr. Regions where SNR$>3$ for the 30\arcsec\ cube are shown with black contours at column density levels of $2^{n}\sigma$ where $n\,=\,0,\,1,\,2,\,3$ and $\sigma\,=\,1.4\times10^{19}\text{cm}^{-2}$ over four channels. The red contours highlight the low column density features at $2^{n}\sigma$ where $n\,=\,-3,\,-2,\,-1$. White dashed contours at [-0.5, -0.25]$\,\times\,10^{19}\,\text{cm}^{-2}$ indicate the \hi\ absorption in the nuclear region of \ngc. The systemic velocity and centre of \ngc\ are indicated by the black dashed horizontal and vertical lines, respectively. The annotations A to F highlight different features discussed throughout the paper.}
\label{pv_0nly}
\end{figure}

We study the channel maps of the 30\arcsec\ data cube, shown in Figure~\ref{chan_maps}, where the tail is visible between 1024 and 1200\,\kms. This velocity range crosses the systemic velocity of the system. To further investigate, we create position velocity diagrams from the 30\arcsec\ data cube across different position angles (PA), as shown in Figure~\ref{pv_0nly}. The cuts are made along the major rotation axis, PA=315\degr, and along the tail feature, PA=348\degr. The black contours corresponds to column densities of $2^{n}\,\sigma$ where $n\,=\,0,\,1,\,2,\,3$, with $\sigma\,=\,1.4\times10^{19}\text{cm}^{-2}$, the average column density at 3~SNR in the 30\arcsec\ cube over four channels of 5.5\,\kms. The red contours at $2^{n}\,\sigma$ where $n\,=\,-3,\,-2,\,-1$, highlight the low column density features. 

The majority of the \hi\ gas is found in the star-forming disk of the galaxy, exhibiting regular rotation as shown by the almost symmetric rotation curves in the position velocity diagram along the major axis. 

From the position velocity slice along the tail, we observe a stream of low column density emission extending beyond the systemic velocity \mbox{($v_{\text{sys}}\,=\,1192$~\kms)} and in the opposite direction to the gas in the \hi\ disk, labelled as feature E. This extended feature corresponds to the northern part of the tail, which implies that the tail is counter-rotating with respect to the \hi\ disk. Additionally, the emission spectrum in Figure~\ref{abs} shows an asymmetry at blue-shifted velocities, which we now attribute to the northern tail feature.

This tail is not the only irregular gas in this galaxy. In the position velocity diagrams, we observe several asymmetries along the rotating disk. On the west side of the disk, between velocities of 1250\,\kms\ and the systemic velocity ($v_{\rm sys}=1192$\,\kms), the \hi\ gas appears to be lagging, producing a beard feature, visible in both position velocity diagrams (feature~C in Figure~\ref{pv_0nly}). This beard feature is most likely caused by \hi\ gas that has been expelled out of the main rotating disk due to galactic winds from supernovae activity \citep[][]{sancisi2008accretion,Fraternali2017review}. 
This expelled \hi\ gas is located out of the plane and is rotating more slowly than the disk \citep{schaap2000vertical}. Further beard effects are seen in feature~A at the blue-shifted velocities (1144 to 1166\,\kms). 

At the edge of the rotation curve, in the blue-shifted velocities, we observe diffuse gas sharply extending towards the systemic velocity, creating a hook effect, feature F, in Figure~\ref{pv_0nly}. Feature F runs almost parallel to the northern tail and is associated with a smaller tail in the south of the disk, as seen in the velocity maps. 
These non-rotating features are discussed in more detail in Section~\ref{disc_ana}.

\subsection{\hi~ Absorption}
\label{hi_abs_sec}

Until now, only \hi\ emission had been detected in \ngc. Thanks to MeerKAT's sensitivity and resolution, we are now able to detect \hi\ absorption against the continuum emission peak ($S_{\text{1.4\,GHz}}\approx18$~mJy/beam) in the nuclear region of \ngc, which was missed in the HIPASS survey.  
The \hi\ absorption is not resolved along the weaker continuum emission of the bar and is detected in the central beam in the highest resolution data cube (8\arcsec). The \hi\ absorption spectrum is shown in the bottom panel of Figure~\ref{abs}, was extracted from the 8\arcsec\ data cube, with the central beam corresponding to a diameter of 664\,pc.
The optical depth, $\tau$, is calculated using the following equation \citep[see for example][]{emonts2010}:
\begin{equation}
    e^{-\tau} = 1 - \frac{S_{\text{abs}}}{S_{\text{cont}}} 
\end{equation}

Where $S_{\text{abs}}$ and $S_{\text{cont}}$ are the absorption and continuum flux, respectively. We are then able to determine the column density of the \hi\ absorption \citep[e.g.][]{morganti2018interstellar}:
\begin{equation}
    N_{\text{\hi}}[\text{cm}^{-2}] = 1.82 \times 10^{18}\,T_{s}[\text{K}] \int \tau(v)~dv~[\text{\kms}] 
\end{equation}

Assuming a spin temperature of $T_s=100$\,K, as commonly done for \hi\ studies \citep[see review from ][]{morganti2018interstellar}.

The peak absorbed flux, determined using a Gaussian fit profile, is $S_{\text{abs}}\approx2\,\text{mJy/beam}$. 
This corresponds to an optical depth of $\tau \sim 0.1$ and a column density of approximately $-1~\times~10^{20}~\text{cm}^{-2}$.
The full width at half maximum (FWHM) of the line is $110$\,\kms\ ($\sim20$~ channels).

In the bottom panel of Figure~\ref{abs}, the absorption spectrum appears symmetric with a single peak that is slightly blue-shifted ($\Delta v \approx $72\,\kms) relative to the systemic velocity of \ngc\ ($v_{\rm sys}=1192$\,\kms), indicated by the vertical black line. The green horizontal dashed lines represent the average noise in the 8\arcsec cube. This absorption is also visible in the position velocity diagrams as feature B in Figure~\ref{pv_0nly}, where its blue-shifted nature is evident. 

The optical depths of the \hi\ absorption line fall within the range typically associated with \hi\ disks \citep[see examples in][]{gallimore1999, maccagni2017abs}. These values are significantly larger than those expected for \hi\ outflows, which typically exhibit optical depths of $\tau\approx0.01$ or less \citep[see][]{morganti2005abs_outflow,gereb2015}. The \hi\ disk traced by the absorption is on the circum-nuclear kiloparsec scale, limited by the spatial resolution of 600~pc, and corresponds to the torus of the AGN which is the region outside the optically thick accretion disk typically found in nearby AGN \citep{Gorkom1989, gallimore1999}.

\section{Discussion}
\label{disc_ana}
In this section, we examine the peculiar \hi\ features revealed by the MeerKAT observations, including the tail, the high-velocity dispersion clouds in the star-forming disk and the \hi\ absorption. We explore the possible origins of these features and their implications for the gas dynamics and evolutionary history of \ngc.

\subsection{3D modelling of the \hi\ disk kinematics}
\label{bbarolo}

To gain further insight into the nature of the extended tail and beards, we analyse the \hi\ kinematics in \ngc\ using tilted ring modelling \citep[e.g.][]{rogstad1974aperture}{}{}. This is performed with \texttt{BBarolo}\footnote{https://editeodoro.github.io/Bbarolo/}, a publicly available software package \citep[][]{teodoro20153d}. 

The fitting was carried out in two stages. First, we defined initial values for the galaxy centre, systemic velocity ($v_{\rm sys}$), rotational velocity ($v_{\rm rot}$), inclination ($i$) and position angle (PA). These were based on the fit parameters from \citet{alonso2018resolving}, who modelled the ALMA molecular gas, as well as parameters from the {\tt SoFiA-2} run. The molecular gas kinematics in the inner 655\,pc of \ngc\ were modelled by \citet{alonso2018resolving} with PA$\,=\,320^{\circ}$, $i\,=\,35^{\circ}$ and $v_{\rm sys}\,=\,1198$\kms\ from \citet{alonso2018resolving}. 
For the initial \hi\ disk model, we used inclination $i\,=\,35^{\circ}$ as a free parameter, while PA\,=\,$308^{\circ}$ and $v_{\rm rot}$\,=\,120\kms were taken from the {\tt SoFiA-2} results. The systemic velocity was fixed at $v_{\rm sys}=1192$\,\kms, also obtained from the {\tt SoFiA-2} run.

\begin{table}
\centering
\caption{Rings comprising the \texttt{Bbarolo} model with \mbox{$v_{\text{sys}}=1192$~\kms} and PA\,=\,315\degr\ for each ring. The table provides the radius (R) in kpc and arcsec, as well as rotational velocity ($v_{\text{rot}}$) and inclination ($i$) for each ring. }
\label{tab:rings}
\begin{tabular}{ccccc}
\hline
 R~[kpc] &  R~[arcsec] &  $v_{\text{rot}}$ [\kms] &  $i$ [deg] & $\text{S}_{\text{int}}$~[mJy\,\kms] \\
\hline

      1 &      15 &      149 &      34 & 26.0   \\
      4 &      45 &      152 &     34 & 37.3\\
      6 &      75 &      162 &     33& 41.5\\
      9 &     105 &      169 &     33& 36.7\\
     11 &     135 &      170 &    32& 30.1\\
     14 &     165 &      173 &     32& 20.1\\
     16 &     195 &      174 &     33& 9.93\\
\hline
\end{tabular}
\end{table}

This process was repeated multiple times for each cube to optimise the model. Some parameters converged to a single value, while others were adjusted from free to fixed, specifically PA\,=\,$315^{\circ}$. 
The models derived from different cubes were consistent with each other. However, since the 30\arcsec\ cube best captures both the asymmetric tails and beards without losing resolution, we utilise this model for further analysis. The final model consists of seven rings, with parameters listed in Table~\ref{tab:rings}.

\begin{figure}
  \centering
   \subfloat{\includegraphics[width=0.5\textwidth, trim=40 60 40 60, clip]{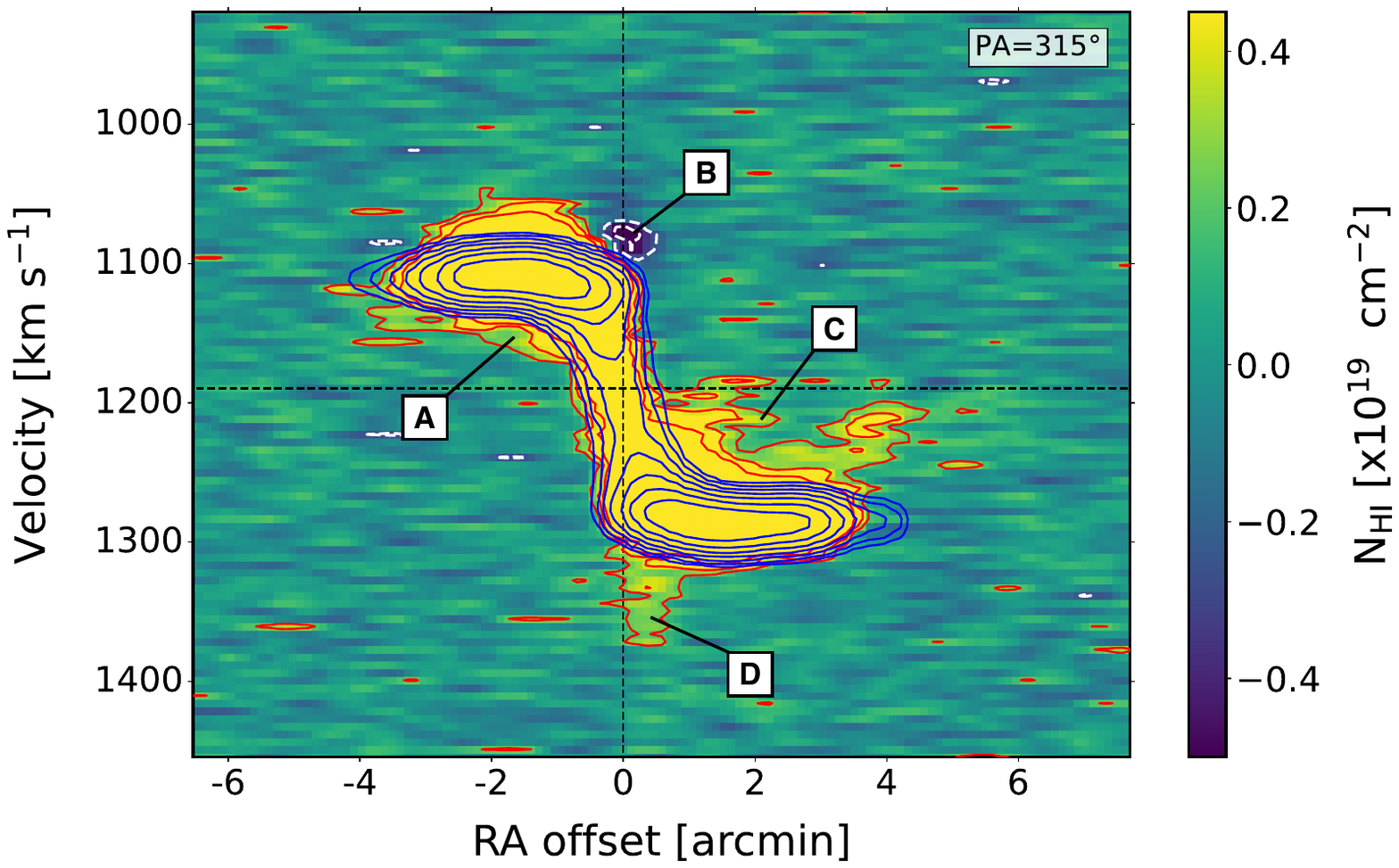}}
  \hfill
   \subfloat{\includegraphics[width=0.5\textwidth, trim=40 60 40 60, clip]{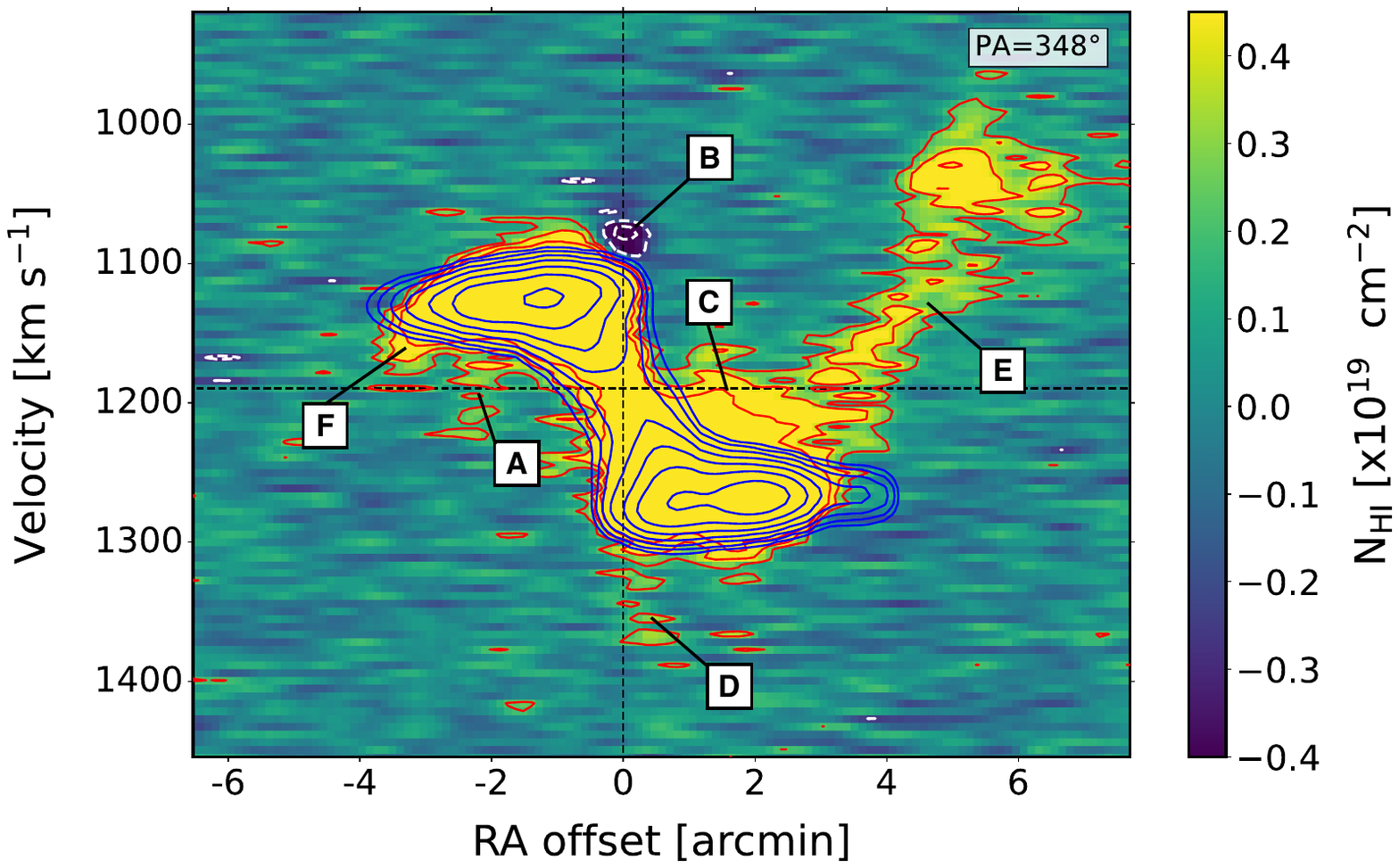}}

\caption[\ngc, Position-velocity diagrams]{\ngc\ position velocity diagrams, similar to  Figure~\ref{pv_0nly}, but with the model in blue contours at column density levels $2^{n}\sigma$ where \mbox{$n\,\in\,(-3,3)$} and \mbox{$\sigma\,=\,1.4\times10^{19}\text{cm}^{-2}$}. 
As in Figure~\ref{pv_0nly}, the low column density red contours are at $2^{n}\,\sigma$, where $n\,=\,-3,\,-2,\,-1$. The annotations A to F highlight different features discussed throughout the paper.}
\label{pv_plots}
\end{figure}

We present the fitted model over the \hi\ emission per velocity channel in Figure~\ref{chan_maps}. 
The model provides a good fit to the \hi\ disk; however, there is a noticeable amount of gas beyond the regions captured by the model, corresponding to features labelled in Figure~\ref{pv_0nly}, suggesting complex gas flows in \ngc. These features are best highlighted in the position–velocity diagrams, where we compare the model (blue contours) with the observed data. Figure~\ref{pv_plots} replicates the position velocity plots from Figure~\ref{pv_0nly} but includes the model for direct comparison.

In Figure~\ref{chan_maps}, between 1145\,\kms\ and 1244\,\kms, we observe gas lagging within the galactic disk. This lagging gas corresponds to the beards (Features A and C) seen in the position velocity diagrams, Figure~\ref{pv_plots}.
The kinematics of the beards (C), described in Section~\ref{hi_emi_sec}, exhibit a significant deviation of $\sim 100$\kms, from the model. This deviation is comparable to the beards observed in NGC~2403, where gas deviates by 60-100\,\kms\ from the main disk \citep{deblok2014N2403}. Another example is NGC\,891, an edge-on galaxy, where \hi\ is found as far as 22~kpc beyond the main disk and lags slightly behind the rotating \hi\ disk \citep{oosterloo2007ngc891}. \citet{oosterloo2007ngc891} suggest that this extended \hi\ halo is primarily due to galactic fountains, though they do not rule out possible contributions from intergalactic accretion in a filament at the outer edges of the disk.

The position velocity diagrams further reveal \hi\ features in \ngc\ that are not easily visible in the channel maps, such as Feature F, B and D. Feature F corresponds to a small southern extension of \hi, also seen in the velocity map (Figure~\ref{mom1_maps}), likely resulting from \hi\ stripping in PGC\,538542. 
Feature B is the \hi\ absorption seen against the continuum sources, while Feature D appears at the corresponding blue-shifted velocities from the \hi\ absorption, suggesting that it may be \hi\ located behind the continuum source.

Between 1024\,\kms and $\sim$1200\,\kms, \hi\ extends northward beyond the disk of \ngc, corresponding to the tail (feature E) in Figure~\ref{pv_plots} and Figure~\ref{mom1_maps}. 
The tail feature exhibits significant asymmetry relative to the disk, with its associated \hi\ gas counter-rotating and extending beyond the galaxy's systemic velocity.
The model allows us to predict the expected location of the gas in the outermost ring. However, the tail's velocity exceeds the expected rotational velocity by $\sim250$\,\kms, further emphasising the asymmetries disk's edge. Given this significant velocity deviation, it is unlikely that the cause of the beards linked to the cause of the tail feature.

To determine the mass of the tail feature, we subtracted the model from the data cube, producing a residual cube (shown as red contours in Figure~\ref{chan_maps}) and a corresponding residual intensity map. We then applied a $3\sigma$ clip and selected a box region to isolate the northern tail. 
After isolating the tail, we estimated its \hi\ mass at $M_\text{\hi} \approx 4 \times 10^{6}$~\msun\, corresponding to $0.1$ per cent of the disk \hi\ mass.

There are several possible explanations for the origin of this tail feature, including ram pressure stripping, recent interactions, or accretion from the environment.
However, the kinematics of the tail, as seen in the position velocity plots and velocity (Figure~\ref{pv_plots} and Figure~\ref{mom1_maps}), do not match the expected behaviour of ram pressure stripping. \citet{serra2023} provide examples of ram pressure stripped galaxies in the Fornax cluster, where velocities smoothly extend from the galaxy's rotation and are dominated by its motion through the intracluster medium (ICM). In contrast, the tail in \ngc\ exhibits counter-rotation, inconsistent with this mechanism.

To further investigate whether the tail results from gas stripping, we assess whether \ngc\ is \hi\ deficient. Using the equations in \citet{Chung2009} and the mean \hi\ surface density within the stellar disk, we calculate an \hi\ deficiency of $\text{def}_\text{\hi} = $\,$-0.87~\pm~0.13$. According to the classification by \citet{yoon2017history}, \ngc\ falls into class 0, which includes galaxies that show no clear signs of gas stripping by the ICM. Instead, these galaxies are either symmetric and \hi-rich or asymmetric with clear signs of tidal interactions. Therefore, the classification of \ngc\ according to \citet{yoon2017history} and the \hi\ deficiency does not indicate that gas is stripped off the galaxy.

The counter-rotating kinematics of the tail suggest that it may be accreting onto \ngc, from various potential sources, such as a recent merger, a small tidally disrupted merging companion dwarf galaxy, or halo gas accretion from the environment. The tail in the north, along with the smaller tail in the south, may indicate ongoing tidal interaction. In the case of a recent massive merger event, we would expect the stars and gas in the system to be more disrupted and diffuse. However, in the disk of \ngc\ the \hi\ gas is regularly rotating, as seen in Figure~\ref{mom1_maps}, and it follows the model well in the position velocity diagrams. Additionally, the deep optical image does not show stars being tidally stripped from the disk, with only an asymmetry in the upper spiral arms being observed. 

While the asymmetries in \hi\ emission in Figure~\ref{full_fov} may suggest an interaction between \ngc\ and IC\,4444, the two galaxies are spatially (234\,kpc) and spectrally (758\,\kms) distant from each other, and therefore not interacting. When compared to galaxy groups within 3500\,\kms\ catalogue \citep{kourkchi2017groups}, \ngc\ and IC\,4444 are not identified as being part of the same group. Furthermore, there has been no evidence of large-scale filamentary structure prior to the discovery of these six \hi\ sources.

In \ngc\ it would take about two full rotations for the \hi\ in the disk to settle into regular rotation \citep{struve2010centaurus}. The timescale for a full rotation is \mbox{$\sim$ 0.7\,Gyr}, calculated using the \mbox{$R_{\text{\hi}}\,=\,16$\,kpc} and $v_{\text{rot}}\,=\,150$\,\kms. This implies that the gas been settled for $\sim1.4$\,Gyr. This timescale does not align with what we would expect if \ngc\ had recently undergone a major merger event.

In Section~\ref{Full_FOV_sec}, we observed that \ngc\ has more surrounding sources than previously thought, with seven \hi\ sources (including IC\,4444). The closest of these source, ID5, is located 104~kpc away and has an optical counterpart in the VST image. Assuming a typical velocity dispersion of $\sim200$\,\kms\ in groups, with a dynamic mass of $\sim24 \times 10^{12}$~\msun\ \citep[see examples in][]{Khosroshahi2007groups}, the galaxies would have collided around $\sim0.5$\,Gyrs ago, suggesting that no recent major interaction has occurred. 
The interaction timescale of ID5 with \ngc\ is close to the timescale required for the gas in \ngc\ to settle into regular rotation. Therefore, if ID5 is \hi\ deficient, it is plausible that it may have passed close to \ngc\ in a fly-by, losing \hi\ gas that is now falling onto the more massive \ngc.

We also examined the VST image for any dwarf galaxies possibly associated with the northern tail. Due to \ngc\ being close to the Galactic plane, the star density in the field of view is high, limiting the depth of our observations. Nonetheless, the surface brightness limit reached by these observations is comparable to that of the deepest optical surveys, such as VEGAS \citep[$\sim27$~mag/arcsec$^2$,][]{2015vegas} and DESI \citep[$\sim22$~mag/arcsec$^2$,][]{2019desi}.
There are several objects in the tail and edge of the galactic disk that could potentially be galaxies associated with \ngc\ or in the background, but due to uncertainties in their redshifts, we cannot definitively confirm their association with \ngc. 

Another possible interaction scenario is that a nearby dwarf galaxy was completely stripped by \ngc. In groups where recent interactions have occurred on massive galaxies, we expect to observe stellar streams connecting the stripped satellite galaxy and the main galaxy, as seen in the less dense environment of NGC~1316 \citep{iodice2017n1316}. However, the tail region of \ngc\ appears to have a low stellar density and no visible stellar stream, making it unlikely that a disrupted galaxy is present within the tail.

The amount of \hi\ in the tail is $4.6\times10^{6}$\,\msun. If we very conservatively assume that the tidally stripped dwarf had a stellar mass equivalent to the \hi\ in the tail, its corresponding apparent magnitude would be $m\,=\,20$. Depending on its extent, this should be visible in the optical VST image, as discussed in Section~\ref{Full_FOV_sec}. 
The VST image in Figure~\ref{cont}, which includes \hi\ continuum contours, reveals an increased concentration of light in the upper stellar spiral arm. This could indicate the presence of a dwarf galaxy embedded within the stellar disk of \ngc.

Additionally, diffuse \hi\ emission is observed around the disk, including a small tail in the south, which could be linked to tidal features. In contrast, the northern tail is significantly larger ($\sim30$\,kpc), extending well beyond the disk, reaching a distance comparable to the galaxy's diameter. 
Currently, we lack metallicity measurements for the gas in the tail. Without this information, we cannot definitively determine whether the tail consists of pristine halo gas or originates from a stripped dwarf galaxy. If the tail resulted from a disrupted dwarf, we would expect an increased stellar density in this region, accompanied by higher metallicity. Deep photometry from the VST data suggest a possible difference in stellar population in the radii of the tail (3\arcmin\,$-$\,4\arcmin), as also indicted by the stellar population study in that region \citep[][]{hoyt2021stellar}. Given these uncertainties, we cannot rule out the possibility that a small dwarf galaxy is associated with \ngc\ and has been stripped.

The kinematics of the northern tail is unusual. It is counter-rotating with respect to the \hi\ disk and extends beyond the systemic velocity of the galaxy by 200\,\kms. The size and kinematics of the tail, along with the low \hi\ deficiency in \ngc, lead us to believe that the tail is more likely \hi\ associated with gas accreting from the environment. Given the tail's extent and low mass, it is possible that a dwarf galaxy either passed by, losing only its gas, or merged directly, contributing to the brighter features in the spiral arms. This suggests that the galaxy is actively accreting the counter-rotating \hi\ gas, along with other non-rotating \hi\ components in the south and at the galaxy's edge. Such gas inflows have predominantly been observed with molecular gas \citep[see examples in][]{combes2013,combes2014,davies2014} due to the sensitivity limit of previous telescopes. The improved sensitivity and resolution of MeerKAT observations are now beginning to reveal \hi\ accretion in low-power AGN, such as Seyfert galaxies.

\subsection{\hii\ and \hi\ comparison }
\label{sec4:h2}

\begin{figure}
  \centering
   \includegraphics[width=0.48\textwidth, trim=20 10 20 10, clip]{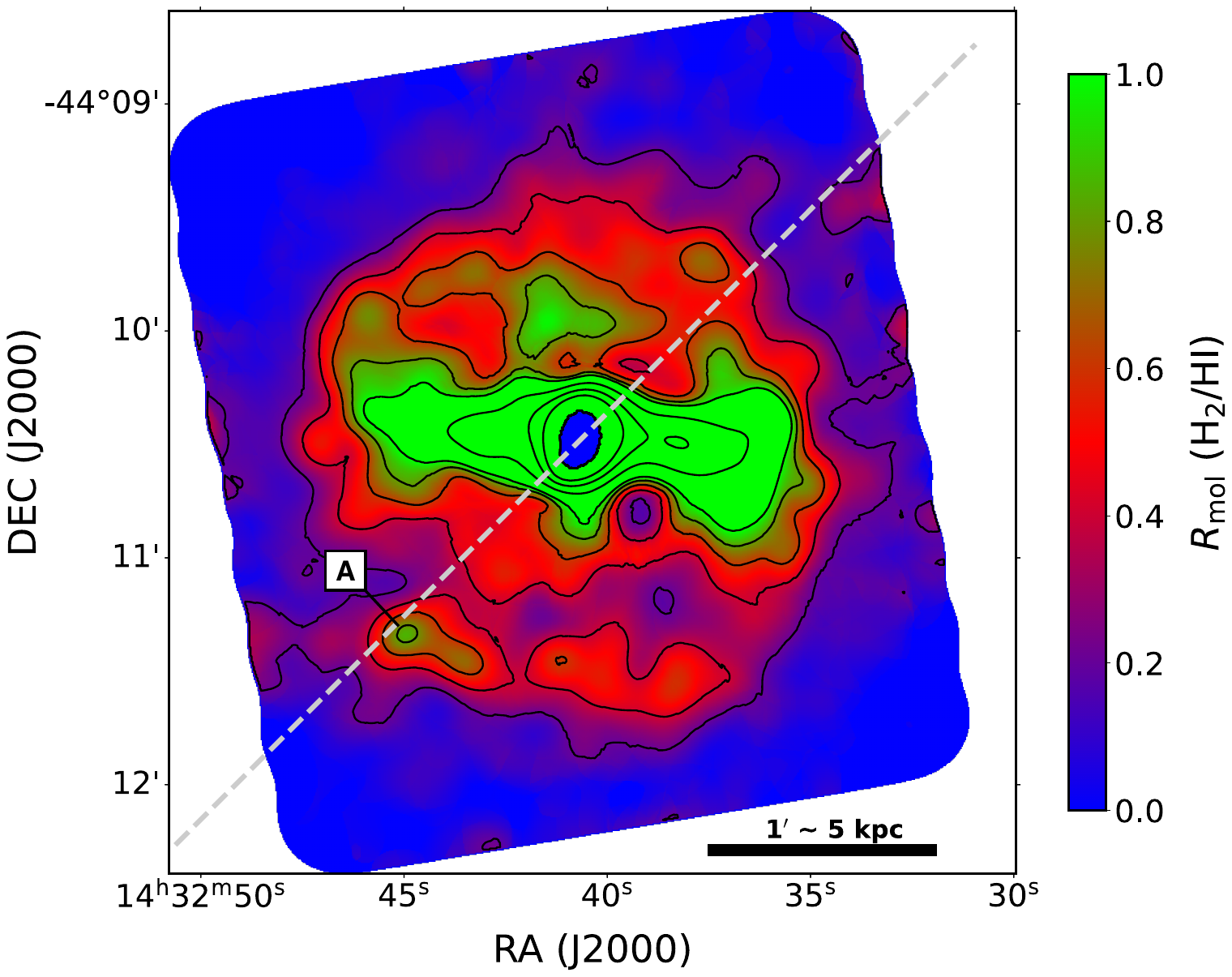}
  
\caption[$R_{\text{mol}}$ comparison ]{The spatial distribution of the \hii$-\text{to}-$\hi\ ratio ($R_{\text{mol}}$) in \ngc\ with contours at [0.2,\,0.4,\,0.6,\,0.8,\,1.0,\,2.0,\,4.0,\,8.0]. The grey dashed line (PA\,=\,315\degr) passes through the centre of \ngc\ and represents the slice used for the position velocity diagram in Figure~\ref{co_pv}. The 8\arcsec\ \hi\ data was smoothed to 11\arcsec\ to match the resolution of the \hii\ data for the $R_{\text{mol}}$ comparison.}
\label{Rmol}
\end{figure}

To further analyse the \hi\ disk, its beards and their connection with star formation, we compare the \hi\ and molecular hydrogen gas (\hii) distributions. We make use of the ALMA CO(2-1) line observations, as mentioned in Section~\ref{anc_data}.
To determine the mass of the \hii\ traced by the CO(2-1), \citet{leroy2021phangs} assumes a standard Milky Way conversion factor of \mbox{$\alpha_{\text{CO}}\,=\,4.35\,$\msun\,$\text{pc}^{-2}$(K\,\kms)$^{-1}$}, which is appropriate for this galaxy type, and CO(2-1)\,-\,to\,-\,CO(1-0) line ratio $\text{R}_{\text{21}}=0.65$. The resulting \hii\ mass is thus calculated to be log($\frac{M_{\text{\hii}}}{\text{\msun}})=9.4$.

We estimate the star formation efficiency (SFE) of \ngc\ using the ratio between \hii\ and \hi\ surface densities, defined as $R_\text{mol}\,=\,\sum_{\text{\hii}}/\sum_{\text{\hi}}$, since \mbox{SFE\,$\propto\,R_{\text{mol}}$} \citep{leroy2008efficiency}.
Using the available multi-wavelength data, we created a map of $R_{\text{mol}}$ (Figure~\ref{Rmol}) by smoothing the 8\arcsec\ MeerKAT intensity map to match the 11\arcsec\ resolution of the ALMA map. Regions where $R_{\text{mol}}<1$ correspond to areas where the \hi\ mass exceeds the \hii\ mass. In contrast, regions with $R_{\text{mol}}\geq1$, where $\text{H}_2$ is dominant, are located along the bar and associated with star-forming regions, as indicated by the continuum image in Figure~\ref{cont}. Along the bar, $R_{\text{mol}}$ remains nearly constant, suggesting efficient star formation.
\citet{leroy2008efficiency} examined the SFE as a function of radius for 23 nearby galaxies and found that nuclear regions exhibit a flat $R_{\text{mol}}$, indicative of efficient star formation. Our results with \ngc\ are consistent with their findings, as expected.

The regions north of the bar exhibit higher $R_{\text{mol}}$ ($R_{\text{mol}}$>0.5) than those south of the bar, except for a few bright spots in the south. These regions, which presumably have higher star formation activity, may be expelling gas out of the disk and could thus be responsible for the beards (A and C) observed in the position velocity diagrams in Figure~\ref{pv_plots}. 

\begin{figure}
  \centering
   \includegraphics[width=0.48\textwidth, trim=40 20 30 20, clip]{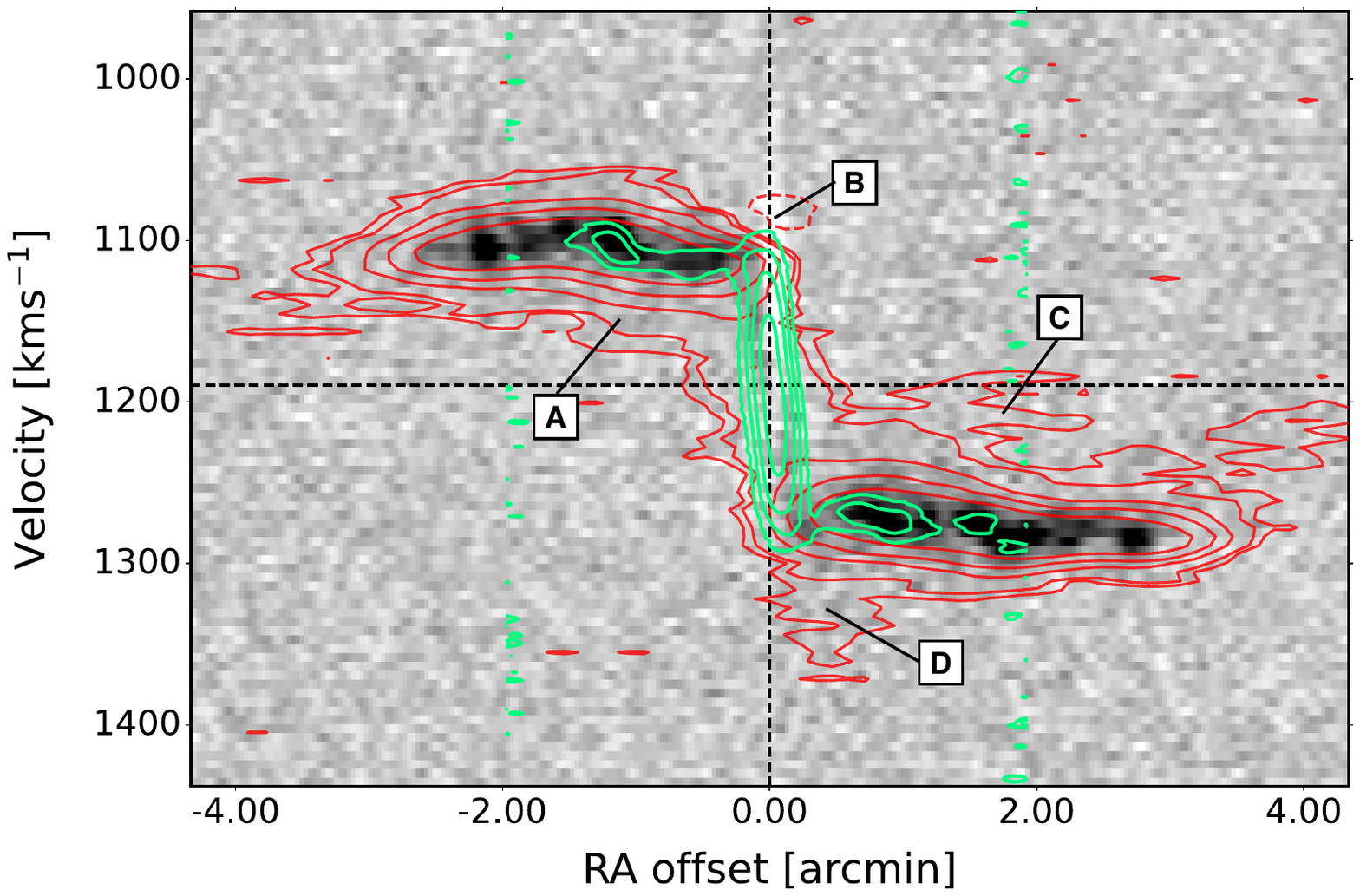}
\caption[\hii\ and \hi\ position velocity comparison ]{This position velocity diagram, slice along PA=315\degr, comparing the \hii\ and \hi\ in inner 13~kpc of \ngc. The position velocity diagram for the 8\arcsec\ cube is shown in greyscale, while the molecular hydrogen (\hii) is shown in green contours at [0.05, 0.125, 0.25, 0.5]~K. The 30\arcsec\ \hi\ cube is represented by red contours at [-0.8, 0.4, 0.8, 2.5, 5.0, 10.0]~mJy/beam. The annotations correspond to the features seen in Figure~\ref{pv_plots}. }
\label{co_pv}
\end{figure}
 
We investigate the kinematics of the southern bright spots, where $R_{\text{mol}}>0.5$, using a position velocity map for the ALMA CO(2-1) PHANGS data. The slice is taken across the centre of \ngc\ at PA\,=\,315\degr, intersecting the left-most bright spot.
The molecular gas position velocity diagram (green) is overlaid on the 8\arcsec\ position velocity diagram (greyscale) at the same slice, shown in Figure~\ref{co_pv}. The red contours represent the 30\arcsec\ position velocity diagram along the same slice.

In the high-resolution (8\arcsec) \hi\ position velocity diagram, we observe that the molecular and neutral atomic phases are linked and exhibit similar kinematics. At $\sim1100$\,\kms, an increase in both \hi\ and \hii\ is associated with the leftmost bright spot and feature A. Given the high spectral resolution of our data, we are able to resolve this elongation, which extends out to 1150\,\kms, confirming that it is \hi\ that has been blown out and is less dense than the \hii. 
These bright spots lie along a spiral arm, where high star formation rates are expected to expel both atomic and molecular gas out of the galactic plane. 
The kinematics associated with the bright spot confirm ongoing star formation and gas recycling in the spiral arms.

Having a complete picture of the molecular and atomic gas phases of the interstellar medium (ISM) in \ngc, we investigate whether \ngc\ possesses a sufficiently large gas reservoir to sustain ongoing star formation.
Star formation efficiency, which is proportional to $R_{\text{mol}}$, is inversely proportional to the depletion time ($\text{t}_{\text{dep}}$); the time required for current star formation to consume the available gas reservoir \citep{leroy2008efficiency}.
Using the \hi\ mass derived from these observations and the \hii\ gas mass from PHANGS, log($\frac{M_{\text{\hii}}}{\text{\msun}})=9.4$, we calculate the \hi\ depletion time to be $\text{t}_{\text{dep}}$(\hi)$\,\approx\,1.79$~Gyrs and the total gas depletion time to be $\text{t}_{\text{dep}}$\,(\hi+\hii)$\,\approx\,3.74$\,Gyrs, using equations from \citet{jaskot2015h}. This short depletion time suggests that the galaxy needs to be replenished with cold gas to sustain the ongoing star formation. The accreting \hi\ tail from the intergalactic medium (IGM), either as a remnant of a tidal interaction or directly from the halo, may contribute to this refuelling.

\subsection{Nuclear region of \ngc}
\label{sec4:abs}

The active circum-nuclear region of \ngc\ has been extensively studied in molecular and ionised gas 
\citep[][]{cresci2015magnum,alonso2018resolving,mingozzi2019magnum,venturi2021magnum}{}{}.
8.4~GHz continuum observations with the VLA by \citet{leipski2006radio} reveal that \ngc\ hosts weakly collimated radio jets, approximately $30$\arcsec\ (2.5\,kpc) in diameter long, extending on both sides of the nucleus along an east-west jet axis. Using the MUSE instrument, \citet{cresci2015magnum} report a bipolar ionisation cone parallel to the jets surrounding the AGN. These ionisation cones extend beyond 1.2\,kpc on either side, aligned with the radio jets. At the centre of these cones, blue-shifted out-flowing ionised gas is observed, with projected velocity $v\sim-450$\,\kms. 
There is also a possibility of a smaller jet structure in the same direction as the outer jets, which may have cleared out the molecular gas from the inner regions. This is suggested by \citet{alonso2018resolving}, who report a region of cleared molecular gas approximately 1\,kpc from the AGN, based on ALMA observations.

As shown in Figure~\ref{abs}, the \hi\ absorption is symmetric and blue-shifted by 72\,\kms. This small shift is unlikely to be tracing the outflow, which in the molecular and ionised phase is observed at $\sim189$~\kms\ and $\sim750$~\kms, respectively \citep{garcia2021multiphase}. 

The absorption feature is also observed in the position velocity diagrams (feature B), where, at the corresponding redshifted velocities, \hi\ appears in emission (feature D). This could trace the gas behind the nucleus, while the absorption traces the gas in front. Given that these two features are seen at symmetric velocities with respect to the systemic velocity, 
it is likely that the absorption is simply tracing the gas in the innermost regions, which are regularly rotating within the bar. The \hi\ tail that is fuelling the rotating disk may also contribute the nuclear activity through the secular motions of the bar. 

\citet{venturi2021magnum} report high \oth velocity dispersion perpendicular to the main radio jets. This perpendicular component is $\sim3$\,kpc in length.
Since the radio jets are low-powered ($\lesssim10^{44} \text{erg s}^{-1}$), \citet{venturi2021magnum} suggest that the jet interacts with the interstellar medium, creating this high-velocity dispersion region perpendicular to the jet. 
We do not see these perpendicular outflows in \hi\ but instead find a deficit of \hi\ in the regions perpendicular to the jets. This also accounts for the increase in $R_\text{mol}$ perpendicular to the bar in the southerly direction, as seen in the $R_\text{mol}$ map in Figure~\ref{Rmol}. This may suggest that the jets are ionising the gas, initially acting on the \hi\ and then destroying the remaining dense molecular clouds. 

VLBI observations will be valuable for understand the nature of the absorption, particularly in determining whether it is associated with jet outflows. Sub-kpc scale angular resolution will enable us to resolve the jets and the \hi\ absorption, providing new insights into the location of this gas within the circum-nuclear regions both spatially and spectrally.

\section{Summary and Conclusions}
\label{conclusion}

We present MeerKAT neutral hydrogen intensity and kinematic maps of the nearby spiral galaxy \ngc\ and its surroundings. 
MeerKAT resolves both diffuse and small-scale structures of \ngc, providing unique insights spanning from circumgalactic to circum-nuclear scales. We also presented VST optical observations and compare the \hi\ distribution with the \hii\ emission from ALMA observations.

Surrounding \ngc\ in the 1.5\,deg$^{2}$ field of view, we identify new \hi\ sources. Excluding IC\,4444, this is the first detection of these sources in \hi\ and, therefore, the first time their spectroscopic redshifts have been determined. We compared the \hi\ sources to the optical VST image and found optical counterparts for ID4, ID5 and ID6. These sources have an \hi\ mass M$_{\text{\hi}}$ $\approx 3\times10^7$\,\msun. The closest source, ID5, has a projected spatial separation of 104\,kpc 
and is too distant to have had a recent major interaction with \ngc\ as the age for the latest possible major interaction is $\sim1.2$\,Gyr. 

While these MeerKAT observations were limited to a single pointing, the discovery of these six lower mass \ngc\ neighbours, combined with previous analyses, suggests this is likely a small group. Future Square Kilometre Array (SKA) observations and wide-field surveys will provide a complete view of these types of gas structures.

In \ngc, the \hi\ emission follows the star-forming spiral arms in a regularly rotating disk with $v_{\text{rot}}=150$~\kms\ and major rotation axis PA=315\degr. 
Given that the disk is regularly rotating to its outermost radius of 16~kpc, the last major interaction that may have unsettled the gas disk kinematics probably occurred $\gtrsim1$~Gyr ago. 
We find small gaseous clouds with low column density ($3\times10^{18}\text{cm}^{-2}$) gas that is associated with \ngc\ but not regularly rotating within the disk. These clouds, seen as lagging gas (beards) in the position velocity diagrams, have likely been expelled from the disk due to ongoing star formation.

The \hii\ and \hi\ comparison of the inner 13\,kpc in \ngc\ reveals regions of increased star formation efficiency, which correspond to the beards in the position velocity diagrams. 
North and south of the bar, there are also regions of increased \hii\ that spatially coincide with the spiral arms, with kinematics matching those of the beards seen in \hi. The \hi\ depletion time calculated from SFE is $\text{t}_{\text{dep}}\text{(\hi)}\approx1.8$\,Gyrs. For the ongoing star formation the cold gas in this galaxy needs to be refuelled by accreting IGM, either from the remnant of a tidal interaction or directly from the halo.

The most noticeable low column density feature in \ngc\ is a diffuse northern tail extending out to 30\,kpc and counter-rotating with respect to the \hi\ disk. This is the first time the tail of diffuse \hi\ ($\sim4\times10^6$~\msun) has been observed to such an extent and resolution. The \hi\ deficiency of \ngc\ is low, $\text{def}_\text{\hi}$\,$=\,-0.87\pm0.13$, 
indicating that the \hi\ is not being stripped from the galaxy \citep{yoon2017history}. The kinematics also do not match what we would expect in a ram pressure stripped galaxy. 

This \hi\ accreting onto the galaxy is either from the halo or due to a dwarf galaxy that has been stripped by \ngc. In the optical VST image of \ngc, this dwarf galaxy is not visible in the \hi\ tail but may be embedded in the face-on stellar disk, as suggested by a bump in one of the upper stellar arms.

The tail's peculiar kinematics, which extends beyond the systemic velocity of the galaxy, suggest that if the tail belonged to a dwarf galaxy, its stellar body (with a mass of $10^4-10^6$~\msun) should have been visible in the deep optical image, since the tail has a \hi\ mass of $\text{M}_{\text{\hi}}\,=\,5.4\times10^6$. More information on the other phases of this gaseous tail and its metallicity would provide new insights into constraining its nature.

\hi\ absorption is detected in \ngc\ against a continuum source. The absorption feature is slightly blue-shifted ($\Delta v \approx -72$\,\kms) with a small \hi\ emission counterpart at higher velocities. This symmetry is due to the bar feature, with gas in front of the continuum source (absorption) and the gas behind the continuum source (\hi\ emission at corresponding, but positive velocity).
The \hi\ tail is fuelling the rotating disk, and through the secular motions of the bar, part of this gas may also fuel the nuclear activity.

We conclude that the neutral hydrogen MeerKAT observation of \ngc\ has resolved an extended northern tail with anomalous kinematics for the first time. Our results demonstrate that this tail represents the accretion of \hi\ onto a regularly rotating \hi\ disk. While we cannot definitively determine the origin of this \hi\ accretion, it may be the result of a stripped dwarf galaxy or simply \hi\ in the environment of \ngc. Ongoing star formation can be observed in both \hi\ emission and in \hi\ to \hii\ comparisons. The \hi\ absorption feature provides insight into the sub-kpc nuclear region of \ngc; however, a more in-depth study of the central 1\,kpc is required to fully understand the physical process occurring around the AGN. 

These new results provide a more holistic view of \ngc. 
MeerKAT observations, such as this one, with high resolution and sensitivity, allow us to gain a more complete understanding of galaxies and how they are formed, evolved and sustained.
As MeerKAT transitions to the SKA, studies such as this will be possible at higher redshifts, making local case studies important examples for broader cosmological research.

\section*{Acknowledgements}
The authors thank the anonymous referee for the useful comments and suggestions that helped improve the paper. 

The authors would also like to thank Dr G. Venturi for the helpful discussions and suggestions.

KC and RPD hereby acknowledge the financial assistance of the South African Radio Astronomy Observatory (SARAO) towards this research (www.sarao.ac.za). RPD's research is funded by the South African Research Chairs Initiative of the Department of Science, Technology, and Innovation; and the National Research Foundation (Grant ID: 77948). 
KC acknowledges financial support from the South African Department of Science and Innovation's National Research Foundation under the ISARP RADIOMAP Joint Research Scheme (DSI-NRF Grant Number 150551).

JH acknowledges support from the UK SKA Regional Centre (UKSRC). The UKSRC is a collaboration between the University of Cambridge, University of Edinburgh, Durham University, University of Hertfordshire, University of Manchester, University College London, and the UKRI Science and Technology Facilities Council (STFC) Scientific Computing at RAL. The UKSRC is supported by funding from the UKRI STFC.

Part of the research activities described in this paper were carried out with contribution of the Next Generation EU funds within the National Recovery and Resilience Plan (PNRR), Mission 4 - Education and Research, Component 2 - From Research to Business (M4C2), Investment Line 3.1 - Strengthening and creation of Research Infrastructures, Project IR0000034 – “STILES - Strengthening the Italian Leadership in ELT and SKA”.

This work has received funding from the European Research Council
(ERC) under the European Union’s Horizon 2020 research and innovation pro-
gramme (grant agreement No 882793 “MeerGas”).

The MeerKAT telescope is operated by the South African Radio Astronomy Observatory, which is a facility of the National Research Foundation, an agency of the Department of Science, Technology, and Innovation. Part of the research is based on data collected with the INAF VST telescope at the ESO Paranal Observatory.

Part of the data published here have been reduced using the CARACal pipeline, partially supported by ERC Starting grant number 679627 “FORNAX”, MAECI Grant Number ZA18GR02, DST-NRF Grant Number 113121 as part of the ISARP Joint Research Scheme, and BMBF project 05A17PC2 for D-MeerKAT. Information about CARACal can be obtained online under the URL: https://caracal.readthedocs.io”. 

We acknowledge the use of the ilifu cloud computing facility - www.ilifu.ac.za, a partnership between the University of Cape Town, the University of the Western Cape, Stellenbosch University, Sol Plaatje University, the Cape Peninsula University of Technology and the South African Radio Astronomy Observatory. The ilifu facility is supported by contributions from the Inter-University Institute for Data Intensive Astronomy (IDIA - a partnership between the University of Cape Town, the University of Pretoria and the University of the Western Cape), the Computational Biology division at UCT and the Data Intensive Research Initiative of South Africa (DIRISA). This work made use of the CARTA (Cube Analysis and Rendering Tool for Astronomy) software (DOI 10.5281/zenodo.3377984 – https://cartavis.github.io).

\section*{Data Availability}
 
The raw visibilities used in this work can be found at the SARAO archive\footnote{https://archive.sarao.ac.za}. The authors may make data products available upon reasonable request.

\bibliographystyle{mnras}
%\bibliography{example} 
\input{mnras_template.bbl}

\newpage
\appendix
\section{\hi\ sources surrounding \ngc}
\label{sip_sec}

\begin{figure*}
\begin{center}
    \includegraphics[width=0.99\textwidth]{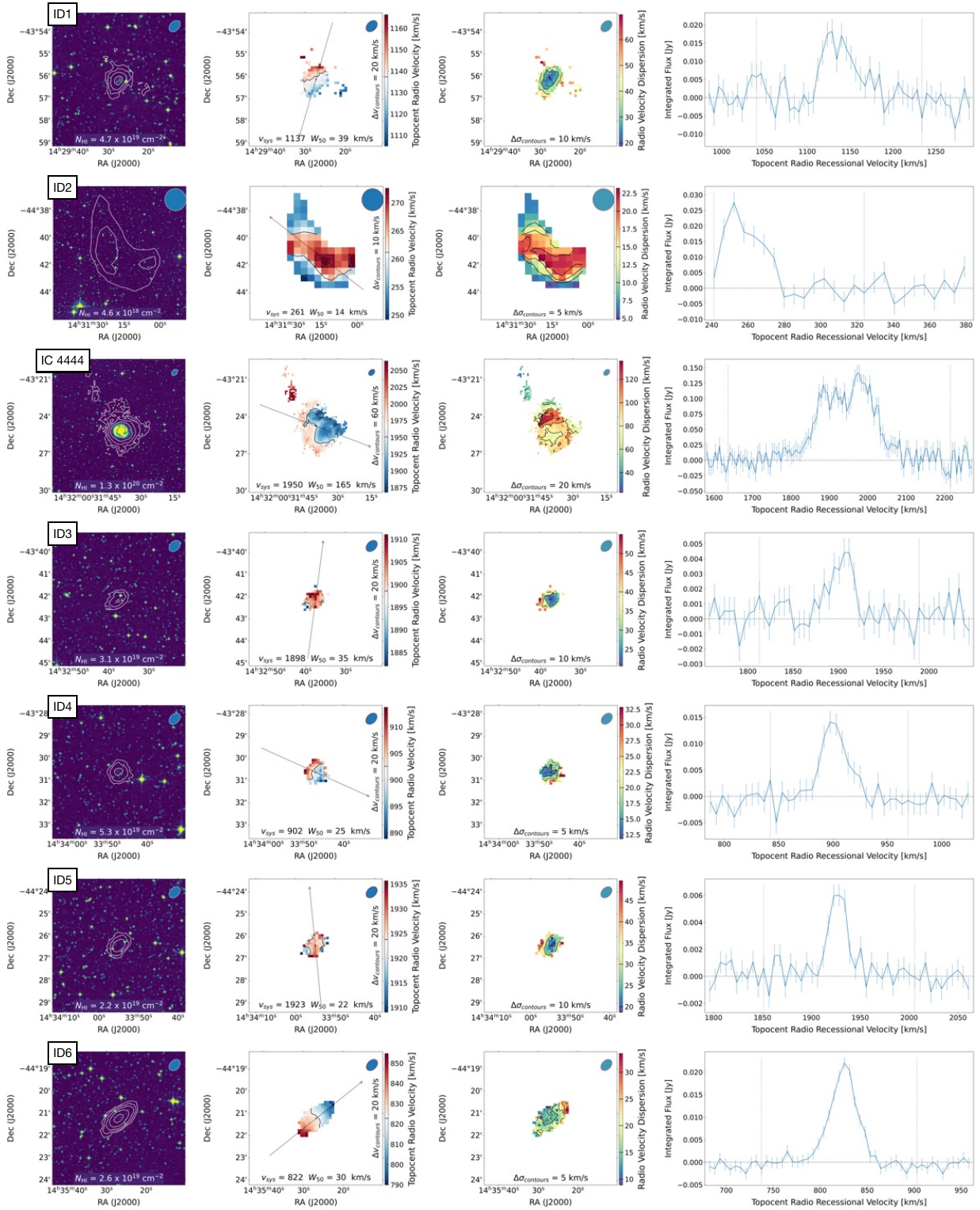}
    \caption[ID plots]{The \hi\ sources surrounding \ngc. From left to right \textit{(i):} Optical Digitized Sky Survey (DSS) images with primary beam-corrected \hi\ column density contours. The contour levels follow $2^{n}\,\times\,N_{\text{\hi}}$\,(with $n=0,1,...$), where the base \hi\ column density values, $N_{\text{\hi}}$, are shown at the bottom of each image. \textit{(ii):} Intensity-weighted velocity field. The arrow indicates the major rotation axis as determined by {\tt SoFiA-2} and the black contour marks the systemic velocity of the source. \textit{(iii):} Velocity dispersion map. \text{(iv):} \hi\ emission spectra with noise. The vertical lines enclose the channel range of the \hi\ emission. For \textit{(i), (ii)} and \textit{(iii)} the contour levels are provided on each image, and the beam shown in the top right. These images are derived from the 30\arcsec\ cube, except for ID2, which is only visible in the 96\arcsec\ cube.
    }
    \label{id_plot}
\end{center} 
\end{figure*}

As mentioned in Section~\ref{obs} there are seven \hi\ sources surrounding \ngc\ (Figure~\ref{full_fov}), six of which are new detection. In this section, we present additional results for these sources. Figure~\ref{id_plot} displays optical images with \hi\ intensity contours, velocity maps, velocity dispersion maps and \hi\ spectra for each source. The contour levels are stated in each image. The arrow on the velocity map indicates the major rotation axis as determined by {\tt SoFiA-2}. These figures were generated using the {\tt SoFiA-2} Imaging Pipeline (SIP)\footnote{https://github.com/kmhess/SoFiA-image-pipeline}. 

\section{Ring properties of \ngc.}

\begin{figure}
\begin{center}
    \includegraphics[width=0.4\textwidth]{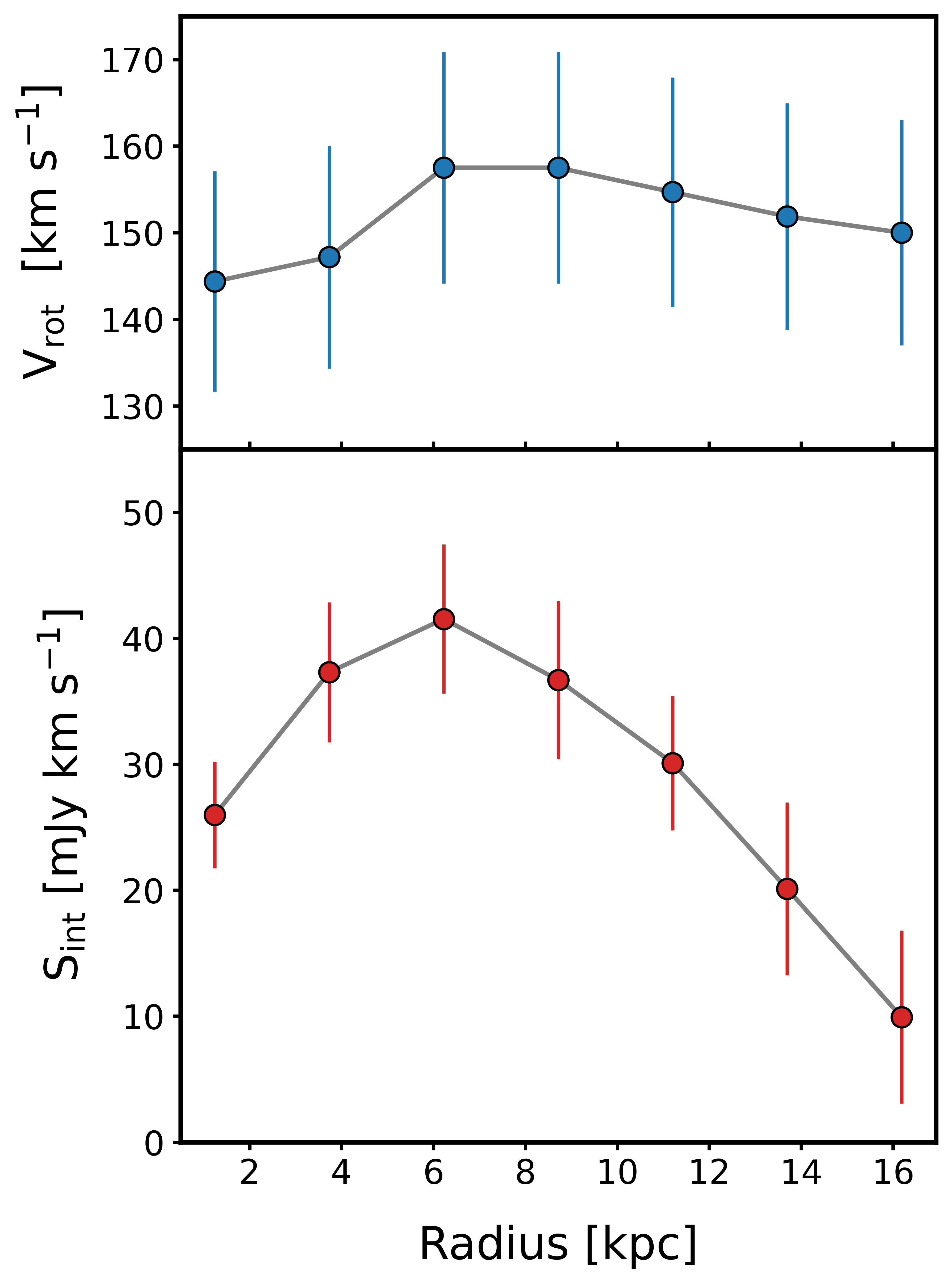}
    \caption{\textit{Top panel:} Rotational velocity plot for each ring in the model. The error bars are 5\,per cent of the value added to the channel resolution of 5.5\,\kms. \textit{Bottom panel:} \hi\ flux density per ring. These results are from the tilted-ring model fitting performed with \texttt{Bbarolo} on the 30\arcsec\ cube. The error bars are the standard deviation values as determined by \texttt{Bbarolo} } 
    \label{b_param}
\end{center} 
\end{figure}

In Section~\ref{bbarolo} we described the generation a model for the \hi\ disk in \ngc\ using \texttt{Bbarolo} with the 30\arcsec\ cube. The rotational velocity ($\text{V}_{\text{rot}}$) was treated as a free parameter when running the \texttt{Bbarolo} fitting. The top panel of Figure~\ref{b_param} shows how this parameter changes with each ring in the model. The errors in $\text{V}_{\text{rot}}$ were calculated by taking 5\,per cent of the value and adding the channel resolution, 5.5\,\kms.

The bottom panel of Figure~\ref{b_param} shows the \hi\ flux density per ring with the standard deviation, as determined by \texttt{Bbarolo}. We observe an increase in flux density up to 7\,kpc, followed by a decreases towards the outer edge of the disk. A decline in \hi\ is noted towards the centre of \ngc, which can be attributed to the \hi\ absorption discussed in Section~\ref{sec4:abs}. The model terminates at the disk's edge and does not account for the northern tail.

\label{lastpage}
\end{document}